\documentclass{jpp}
\usepackage{graphicx}

\usepackage[utf8]{inputenc}
\usepackage[T1]{fontenc}
\usepackage{amsmath}
\usepackage[capitalize]{cleveref}
\usepackage{soul,xcolor}
\usepackage[labelformat=simple]{subcaption}

\setstcolor{red}

\shorttitle{Near-Axis Expansion at Arbitrary Order}
\shortauthor{R. Jorge, W. Sengupta, M. Landreman}

\title{Near-Axis Expansion of Stellarator Equilibrium at Arbitrary Order in the Distance to the Axis}

\author{R. Jorge\aff{1}
  \corresp{\email{rjorge@umd.edu}},
  W. Sengupta\aff{2}
 \and M. Landreman\aff{1}}

\affiliation{\aff{1}Institute for Research in Electronics and Applied Physics, University of Maryland, College
Park, MD 20742, USA
\aff{2}Courant Institute of Mathematical Sciences, New York University, New York NY 10012, USA}

\begin{document}

\maketitle
\begin{abstract}
A direct construction of equilibrium magnetic fields with toroidal topology at arbitrary order in the distance from the magnetic axis is carried out, yielding an analytical framework able to explore the landscape of possible magnetic flux surfaces in the vicinity of the axis. This framework can provide meaningful analytical insight on the character of high-aspect-ratio stellarator shapes, such as the dependence of the rotational transform and the plasma beta-limit on geometrical properties of the resulting flux surfaces. The approach developed here is based on an asymptotic expansion on the inverse aspect-ratio of the ideal MHD equation. The analysis is simplified by using an orthogonal coordinate system relative to the Frenet-Serret frame at the magnetic axis. The magnetic field vector, the toroidal magnetic flux, the current density, the field line label and the rotational transform are derived at arbitrary order in the expansion parameter. Moreover, a comparison with a near-axis expansion formalism employing an inverse coordinate method based on Boozer coordinates (the so-called Garren-Boozer construction) is made, where both methods are shown to agree at lowest order. Finally, as a practical example, a numerical solution using a W7-X equilibrium is presented, and a comparison between the lowest order solution and the W7-X magnetic field is performed.
\end{abstract}

\section{Introduction}

To successfully confine a high-temperature plasma, with the ultimate goal of yielding a net energy gain from the resulting nuclear fusion reactions, the plasma pressure and electromagnetic forces should be balanced for sufficiently long periods of time.
For this reason, the study of plasma equilibria lays at the fundamental level of magnetic confinement studies.
Such magnetic fields are found by solving the ideal magnetohydrodynamics (MHD) equation
\begin{equation}
    \mathbf J \times \mathbf B = \nabla p.
\label{eq:idealMHD}
\end{equation}
In this work, we focus on obtaining three-dimensional magnetic equilibrium fields suitable for fusion devices, such as tokamaks and stellarators.
Compared to tokamaks, stellarators have the advantage of eliminating instabilities and difficulties related to current-driven modes of operation.
However, the degrees of freedom related to the solution of \cref{eq:idealMHD} in a non-axisymmetric geometry increases substantially [approximately one order of magnitude \citep{Boozer2015}] compared to its axisymmetric counterpart.
An outstanding challenge is therefore to understand the landscape of three-dimensional equilibrium magnetic fields and to identify its most relevant cases for the success of the fusion program.

The current effort of stellarator optimization is to find shapes of external coils and currents that yield equilibrium magnetic fields with good plasma confinement.
Such optimization effort has lead to breakthroughs related to stellarator confinement, namely in the field of neoclassical transport (such as quasisymmetry), MHD stability and turbulence associated with drift waves \citep{Grieger1992, Mynick2006, Mynick2010}.
These efforts rely mainly on computational tools that provide solutions that are largely dependent on the initial point used in configuration space, where such dependencies are usually unknown.
In this work, we perform a theoretical construction of stellarator equilibrium fields that can act as a guideline for the computational stellarator optimization program {by providing a practical tool that can generate good initial points for conventional optimization algorithms} while also allowing the theoretical analysis of their confinement properties independently of the {chosen} algorithms.

The near-axis framework is based on an asymptotic expansion of the equilibrium fields in powers of the inverse aspect ratio $\epsilon$, where
\begin{equation}
    \epsilon = \frac{a}{R} \ll 1,
\end{equation}
with $a$ the maximum perpendicular distance from the axis to the plasma boundary and $R$ the minimum of the local radius of curvature of the magnetic axis.
When solving \cref{eq:idealMHD}, we focus on a system where the plasma $\beta$ is small, i.e.,
\begin{equation}
    \beta = \frac{\overline p}{\overline B^2/8\pi} \ll 1,
\label{eq:betaexp}
\end{equation}
where $\overline p$ and $\overline B$ in \cref{eq:betaexp} are taken to be the  pressure and average magnetic field strength on axis.
Finally, the magnetic field $\mathbf B$ is written in terms of the toroidal magnetic flux $\psi$ and a field line label $\alpha$ using the Clebsch representation
\begin{equation}
    \mathbf B = \nabla \psi \times \nabla \alpha,
\label{eq:clebsch}
\end{equation}
which is a way of {locally} writing divergence-free vector fields \citep{Helander2014}.
Within the near-axis formalism, the fields $\mathbf B, \psi$ and $\alpha$ are expanded in a power series in $\epsilon \rho/a$, with the $\rho$ distance from the magnetic axis to an arbitrary point along the plane locally perpendicular to the axis.

The construction of magnetic field equilibria using a near-axis framework has mainly followed one of two approaches, namely using a direct or an inverse coordinates approach.
In the direct method, pioneered by Mercier, Solov'ev and Shafranov \citep{Mercier1964, Solovev1970}, the magnetic flux surface function $\psi$ is found explicitly in terms of the Mercier coordinates $(\rho, \theta, s)$, with $\theta$ the angular polar coordinate in the plane locally perpendicular to the magnetic axis and $s$ the arclength function of the magnetic axis curve.
This method allowed for several significant analytical results in the context of stellarator equilibria.
An estimate for equilibrium and stability $\beta$-values using the direct approach was first given in \citet{Lortz1976, Lortz1977} by carrying out the expansion up to third order in $\rho$.
Higher order formulations of the direct approach were also used in \citet{Bernardin1986,Salat1995} to prove important geometric properties of MHD equilibria and, more recently, in \citet{Chu2019}, to obtain a generalized Grad-Shafranov equation for near-axis equilibria with constant axis curvature.
Finally, we note that the direct method can also be used to derive a Hamiltonian formulation for the magnetic field lines and obtain adiabatic invariants to successively higher-order in $\rho$ \citep{Bernardin1985}.

In contrast, in the inverse method, the spatial position vector $\mathbf r$ is obtained as a function of magnetic coordinates involving $\psi$, the toroidal magnetic flux, such as Hamada or Boozer coordinates.
The first use of the inverse method can be traced back to the work of Lortz and Nuhrenberg [see Appendix II of \citet{Lortz1976}], where a near-axis expansion in Hamada coordinates was related to an expansion in the direct method to evaluate the Mercier stability criterion.
An inverse coordinate description relying solely on  Boozer coordinates was pioneered by  \citet{Garren1991a, Garren1991}, further extended to allow vanishing curvature and the use of standard cylindrical coordinates in \citet{Landreman2018, Landreman2019a}.
Boozer coordinates have the advantage that the particle guiding center drift-trajectories are determined by the magnetic field strength $B$ only [in contrast with the magnetic field vector $\mathbf B(\mathbf r)$] \citep{Boozer1981}.
Furthermore, the Garren-Boozer construction allows for a practical procedure to directly construct MHD equilibria optimized for neoclassical transport without numerical optimization, i.e., to obtain analytical quasisymmetric fields, while showing that at third order in $\sqrt{\psi}$ the requirement of quasisymmetry leads to an overdetermined system of equations \citep{Garren1991}.
To lowest order, however, it was shown that the core shape and rotational transform of many optimization-based experimental devices could be accurately described by the Garren-Boozer construction \citep{Landreman2019}, showing that a near-axis framework can potentially be used as an accurate analytical model for modern stellarator configurations.
%

In this work, for the first time, the direct method is formulated at arbitrary order in $\epsilon$ (and hence $\rho$) for both vacuum and finite-$\beta$ systems.
The use of the direct method has several advantages with respect to the inverse one, which are explored here.
First, while the inverse approach relies on the existence of a flux surface function $\psi$ to define its coordinate system, the direct method allows for the construction of magnetic fields with resonant surfaces (such as magnetic islands) and can provide analytical constraints for the existence of magnetic surfaces.
Second, in the direct method, the magnetic axis is defined in terms of the vacuum magnetic field, allowing the determination of a Shafranov shift and plasma $\beta$ limits when MHD finite $\beta$ effects are included.
Finally, due to the use of an orthogonal coordinate system, the algebra is simplified considerably, allowing for the determination of the asymptotic expansion of $\mathbf B$, $\psi$, $\alpha$ and $\iota$ at arbitrary order in $\epsilon$.

This paper is organized as follows.
In \cref{sec:nearaxis} the near-axis framework is introduced, focusing on the construction of an orthogonal coordinate system based on the magnetic axis and the asymptotic expansion of the physical quantities of interest.
The asymptotic expansions of the vacuum magnetic field $\mathbf B$, the magnetic flux surface function $\psi$, and the magnetic field line label $\alpha$ are obtained in terms of Mercier coordinates in \cref{sec:vacuum}, while the finite $\beta$ case is presented in \cref{sec:mhd}.
In particular, the lowest order vacuum solutions are cast in terms of geometrical quantities of the elliptical flux surface, such as the eccentricity and rotation angle.
In \cref{sec:rotationaltransform}, the rotational transform $\iota$ is computed based on the solution for $\alpha$, and the rotational transform on axis is analytically evaluated and interpreted based on geometrical considerations.
A comparison with an indirect method, particularly with the Garren-Boozer construction, is performed in \cref{sec:comparisongb} where equivalence between both approaches is shown at lowest order.
Finally, a numerical solution of the lowest order system of equations is obtained in \cref{sec:numerical} by comparing with a W7-X equilibrium profile.
The conclusions follow.

\section{Near-Axis Framework in Mercier's Coordinates}
\label{sec:nearaxis}

\subsection{Mercier's Coordinate System}

\begin{figure}
    \centering
    \includegraphics[width=0.8\textwidth]{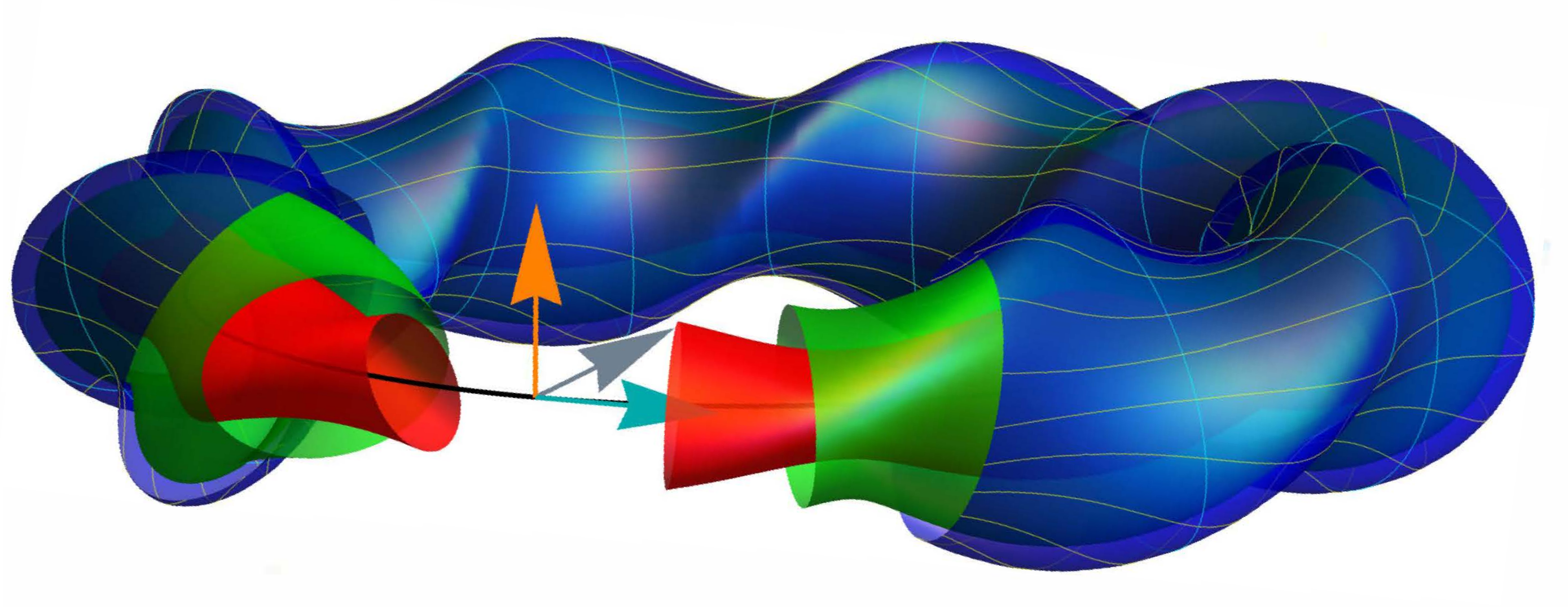}
    \caption{Example of a magnetic surface using Mercier's coordinates, where the radial coordinate $\rho$ is prescribed as a function of $s$ and $\omega=\theta+\gamma$ in order to obtain a toroidal shape with elliptical cross-section. The magnetic axis curve is shown in black, together with the Frenet-Serret unit vectors, namely the tangent (cyan), normal (gray) and binormal (orange) unit vectors. The blue and yellow lines in the outermost surface denote the curves with constant $\theta$ and $s$, respectfully.
    }
    \label{fig:fsexample}
\end{figure}

In this section, leveraging the work in \citep{Mercier1964, Solovev1970}, we construct an orthogonal coordinate system associated with a particular field line of force $\mathbf r_0(s)$, which is taken to be the magnetic axis curve.
We let $L$ denote the total length of $\mathbf r_0$, such that $0 \le s \le L$.
The unit tangent vector $\mathbf t$ is defined as
\begin{equation}
    \mathbf t=\mathbf r_0'(s),~0 \le s \le L.
\label{eq:unitangent}
\end{equation}
Using the fact that $\mathbf t'(s)$ is orthogonal to $\mathbf t$, the unit normal vector $\mathbf n$ is defined as $\mathbf n=\mathbf t'(s)/\kappa$ with $\kappa=|\mathbf t'(s)|=|(\mathbf t \cdot \nabla) \mathbf t|$ the curvature, while the unit binormal vector $\mathbf n$ obeys $\mathbf b = \mathbf t \times \mathbf n$.
The triad $(\mathbf t, \mathbf n, \mathbf b)$ forms a right-handed system of orthogonal unit vectors, usually called the Frenet-Serret frame \citep{Spivak1999}, which obey the following set of first-order differential equations
\begin{align}
    \mathbf t'(s)&= \kappa \mathbf n,\label{eq:FS1}\\
    \mathbf n'(s)&=-\kappa \mathbf t + \tau \mathbf b,\label{eq:FS2}\\
    \mathbf b'(s)&=-\tau \mathbf n,\label{eq:FS3}
\end{align}
with $\tau$ the torsion.
Explicit expressions for the curvature and torsion when the curve $\mathbf r_0$ is parametrized in terms of a parameter $t$ other than the arclength $s$ (e.g., the toroidal angle $\Phi$ in cylindrical coordinates) can be obtained using
\begin{equation}
    \kappa(t)=\frac{|\mathbf r'(t)\times \mathbf r''(t)|}{|\mathbf r'(t)|^3},
\label{eq:curvfs}
\end{equation}
and
\begin{equation}
    \tau(t)=\frac{\left(\mathbf r'(t) \times \mathbf r''(t)\right)\cdot \mathbf r'''(t)}{|\mathbf r'(t) \times \mathbf r''(t)|^2}.
\label{eq:torfs}
\end{equation}
It can be shown that any curve (such as the magnetic axis) can be described only with $\kappa$ and $\tau$ [see, e.g., \citep{Spivak1999}].
Given $\kappa$ and $\tau$, the Frenet-Serret frame can then be found using \cref{eq:FS1,eq:FS2,eq:FS3} and a set of initial conditions.
An example of a magnetic axis curve is shown in \cref{fig:fsexample}, together with the Frenet-Serret unit vectors, namely the tangent (blue), normal (green) and binormal (red) unit vectors.

We now let $\rho$ denote the distance between an arbitrary point $\mathbf r$ to the axis $\mathbf r_0$ in the plane $(\mathbf n, \mathbf b)$ orthogonal to the axis, and let $\theta$ be the angle measured from the normal to $\mathbf r-\mathbf r_0$.
The radius vector $\mathbf r$ can then be written as
\begin{equation}
    \mathbf r = \mathbf r_0(s) + \rho \cos \theta \mathbf n(s) + \rho \sin \theta \mathbf b(s).
\label{eq:posMercier}
\end{equation}
We note that the set of coordinates $(\rho, \theta, s)$ can be made orthogonal by introducing the angle $\omega$, defined by
\begin{equation}
    \omega = \theta + \gamma(s),~\gamma(s)=\int_0^s \tau(s') ds',
\end{equation}
with $\gamma$ the integrated torsion.
An example of a magnetic surface constructed using Mercier's formalism is shown in \cref{fig:fsexample}, where the radial coordinate $\rho$ was chosen to be a function of $s$ and $\omega$ in order to obtain poloidal cross-sections with an elliptical shape.
The explicit expression of $\rho(s,\omega)$ for the case of an elliptical expression can be found in \cref{sec:vacuum}.
The determinant of the metric tensor is given by 
\begin{equation}
    \sqrt{g} = \rho h_s
\label{eq:sqrtg}
\end{equation}
in both the $(\rho,\theta,s)$ and $(\rho,\omega,s)$ coordinate systems.
However, the metric tensor $g_{ij}$ is diagonal only when expressed in terms of the $(\rho,\omega,s)$ coordinates, with $g_{ij}=\text{diag}(1,\rho^2,h_s^2)$.
In \cref{eq:sqrtg}, we defined $h_s$ as
\begin{equation}
    h_s=1-\kappa \rho \cos {(\omega-\gamma)}.
\end{equation}
{In the following, we assume that the plasma boundary is close enough to the axis such that $\rho<1/\kappa$ leading to a non-zero Jacobian across the whole plasma volume.}
Finally, we introduce a Cartesian coordinate system in the $(\mathbf n, \mathbf b)$ plane by defining $x=\rho \cos \theta$ and $y=\rho \sin \theta$, yielding
\begin{equation}
    \mathbf r = \mathbf r_0(s) + x \mathbf n(s) + y \mathbf b(s).
\label{eq:poscart}
\end{equation}
We note that the position vector in \cref{eq:poscart} coincides with the definition in the inverse coordinate method \citep{Garren1991a} only to lowest order in $\epsilon$.
A key difference between the direct method pursued here and the inverse method in \citep{Garren1991a} is that the latter includes a finite contribution in the $\mathbf t$ direction in \cref{eq:poscart}, leading to distinct vectors $\mathbf r - \mathbf r_0(s)$ between both the direct and the inverse method at the same point $\mathbf r$.

\subsection{Power Series Expansion}

In this section, we show how to construct the asymptotic series related to a physical quantity {f} in terms of $\epsilon$ and $\rho$.
{As shown below, such construction imposes conditions on the series coefficients of $f$. In \cref{sec:vacuum}, we prove that the magnetic field satisfies such conditions in the vacuum case at all orders.}
In what follows, we normalize $\rho, x$ and $y$ to the characteristic perpendicular scale $a$ and the quantities $\kappa, \tau$ and $s$ to the characteristic parallel scale $R$.
The magnetic field is normalized to a constant $\overline{B}=\int_0^L B_0(s)ds/L$, with $B_0(s)$ the magnetic field on-axis and $\psi$ is normalized to $\overline B R^2$.
{The asymptotic expansion of a function $f$ is then constructed by noting that any analytic function $f$ has a Taylor expansion near the origin $(x,y)=(0,0)$ of the form}
\begin{equation}
    f(x,y,s)=\sum_{p=0}^\infty\sum_{j=0}^\infty f_{pj}(s)x^p y^j \epsilon^{p+j}.
\label{eq:phixys}
\end{equation}
Similarly, using \cref{eq:poscart}, we {write} $f$ {as}
\begin{equation}
    f(\rho,\omega,s)=\sum_{n=0}^\infty f_n(\omega,s)\epsilon^n \rho^n.
\label{eq:phirhoomega}
\end{equation}
The equality between the two expansions in \cref{eq:phirhoomega,eq:phixys} yields
\begin{equation}
    f_n(\omega,s)=\sum_{p=0}^n f_{pn-p}(s)\cos^p (\theta) \sin^{(n-p)} (\theta){=\sum_{p=0}^n f_{np}^c(s) \cos{(p \theta)}+f_{np}^s(s)\sin{(p \theta)}}.
\label{eq:analyticity}
\end{equation}
Equation (\ref{eq:analyticity}) shows that $f_n$ can be written as a Fourier series in terms of $\cos p \theta$ and $\sin p \theta$ with only even or odd values of $p$ in the range $0\le p \le n$ depending on whether $n$ is even or odd \citep{Kuo-Petravic1987}.
{This is proven} in Appendix \ref{app:Fourierphi}.
{In the following, when dealing with Mercier's coordinates $(\rho,\omega,s)$, analyticity is shown in the form of \cref{eq:analyticity} rather than \cref{eq:phixys}}

We expand the components of the magnetic field $\mathbf B$ as $\mathbf B=B_\rho \mathbf e_\rho + B_\omega \mathbf e_\omega + B_s \mathbf e_s$ with $\mathbf e_\rho=\cos \theta \mathbf n + \sin \theta \mathbf b$, $\mathbf e_\omega=-\sin \theta \mathbf n + \cos \theta \mathbf b$ and $\mathbf e_s= \mathbf t$ using \cref{eq:phirhoomega}, i.e.,
\begin{align}
    B_\rho&=\sum_{n=0}^{\infty} B_{\rho n}(\omega,s) \epsilon^n \rho^n,
\end{align}
and similarly for $B_\omega, B_s$ and $\psi$.
The fields $\mathbf B, \psi$ and $\alpha$ are split into a vacuum and a finite $\beta$ component as 
\begin{equation}
    \mathbf B = \mathbf B^0 + \mathbf B^1,
\label{eq:Bexpansionbeta}
\end{equation}
with $\mathbf B^0$ the vacuum magnetic field in the absence of a plasma and $\mathbf B^1$ a linear perturbation in $\beta$.
A similar split is applied to $\psi$ and $\alpha$.
The linear perturbations satisfy
\begin{equation}
    \left|\frac{\mathbf B^1}{\mathbf B^0}\right|\sim \frac{\psi^1}{\psi^0}\sim\frac{\alpha^1}{\alpha^0}\sim \beta \ll 1
\end{equation}
We first consider the vacuum case and obtain a set of equations for the coefficients $B^{0}_{\rho n}, B^{0}_{\omega n}, B^{0}_{s n}, \psi^0_n$ and $\alpha^0_n$.

\section{Vacuum Configuration}
\label{sec:vacuum}

In vacuum, the magnetic field is irrotational, i.e., $\nabla \times \mathbf B=0$.
We therefore define the magnetic scalar potential $\phi$ as
\begin{equation}
    \mathbf B^0 = \nabla \phi,
\label{eq:B0vac}
\end{equation}
with $\phi$ normalized to $\overline B R$ and $\nabla$ normalized to $R$.
Using \cref{eq:B0vac} and $\nabla \cdot \mathbf B = 0$, we find that $\phi$ satisfies Laplace's equation, $\nabla^2 \phi=0$.
In normalized Mercier coordinates $(\rho,\omega,s)$, the gradient operator $\nabla \phi$ can be written as 
\begin{equation}
    \nabla \phi = \frac{1}{\epsilon}\frac{\partial \phi}{\partial \rho}\mathbf e_\rho+\frac{1}{\epsilon \rho}\frac{\partial \phi}{\partial \omega}\mathbf e_\omega+\frac{1}{h_s}\frac{\partial \phi}{\partial s}\mathbf e_s,
\label{eq:gradientop}
\end{equation}
while Laplace's equation reads
\begin{equation}
    \frac{1}{\rho}\frac{\partial}{\partial \rho}\left(h_s \rho \frac{\partial \phi}{\partial \rho}\right)+\frac{1}{\rho^2}\frac{\partial}{\partial \omega}\left(h_s\frac{\partial \phi}{\partial \omega}\right)+\epsilon^2\frac{\partial}{\partial s}\left(\frac{1}{h_s}\frac{\partial \phi}{\partial s}\right)=0.
\label{eq:laplacephi}
\end{equation}

As $\mathbf r_0$ is the axis of the vacuum magnetic field, $\mathbf B^0$, we impose both the radial and angular vacuum magnetic field to vanish when $\rho = 0$, and the magnetic field on-axis $B_0$ to be a function of $s$ only, i.e., $\mathbf B^0(\rho=0)=B_0(s)\mathbf e_s$.
This sets $B^0_{\omega 0}=B^0_{\rho 0}=0$ and $B^0_{s0}=B_0(s)$.
The lowest order solution of \cref{eq:B0vac} is then given by
\begin{equation}
    \nabla \phi = B_0(s)\mathbf e_s + O(\epsilon),
\end{equation}
which, up to $O(\epsilon)$, yields
\begin{equation}
    \phi=\int_0^s B_0(s')ds'+O(\epsilon^2).
\end{equation}

\subsection{Vacuum Magnetic Scalar Potential}

We now expand $\phi$ using \cref{eq:phirhoomega} and collect terms of the same order in $\epsilon \rho$ in \cref{eq:laplacephi} in order to obtain a single equation for $\phi_n$.
In the following, we define $\dot \phi=\partial_\omega \phi$ and $\phi'=\partial_s \phi$ for the partial derivatives with respect to $\omega$ and $s$, respectively.
We first focus on the $O(\epsilon^2)$ term of \cref{eq:laplacephi}
\begin{equation}
    \ddot \phi_2 + 4 \phi_2 = - B_0'(s).
\label{eq:phi2eq}
\end{equation}
The homogeneous solution of \cref{eq:phi2eq} can be written as a linear combination of $\sin 2 \omega$ and $\cos 2 \omega$ terms, while the particular solution is given by $-B_0’/4${, i.e.
\begin{equation}
	\phi_2 = -\frac{B_0'}{4}+c_1 \sin 2 \omega + c_2 \cos 2 \omega.
\end{equation}
}
{For convenience, and} to later obtain a direct relation between the coefficients of $\sin 2 \omega$ and $\cos 2 \omega$ and the geometric parameters of the flux surface function, we introduce the functions $\delta(s), \eta(s)$ and $\mu(s)=\tanh \eta(s)$, and write $\phi_2$ as
\begin{equation}
    \phi_2=\frac{B_0}{2}\left[(\ln B_0^{-1/2})'+\mu u' \sin 2u-\frac{\eta'}{2}\cos 2 u\right],~u=\omega - \gamma(s)+\delta(s)
\label{eq:phi2soldeltaeta}
\end{equation}
The integration constants $\delta(s)$ and $\eta(s)$ that characterize the $O(\epsilon^2)$ scalar potential $\phi$ are arbitrary (periodic) functions of $s$ can be used to impose additional constraints on the magnetic field $\mathbf B$, such as quasisymmetry or omnigeneity.

Focusing on the $O(\epsilon^3)$ term of \cref{eq:laplacephi}, we obtain the following equation for $\phi_3$
\begin{equation}
    9 \phi_3+\ddot \phi_3=-B_0 \cos \theta  \kappa'+\kappa\left[2  \phi_{22}^c \cos (\theta+2 \delta )+2 \phi_{22}^s \sin (\theta +2 \delta)-B_0 \tau  \sin \theta -2 B_0' \cos \theta\right],
\label{eq:phi3eq}
\end{equation}
with $\phi_{22}^c = - B_0 \eta'/4$ and $\phi_{22}^s = B_0\mu u'/2$.
Similarly to \cref{eq:phi2soldeltaeta}, we write $\phi_3$ as
\begin{equation}
    \phi_3=\phi_{31}^c \cos u + \phi_{31}^s \sin u + \phi_{33}^c \cos 3u + \phi_{33}^s \sin 3u.
\label{eq:solphi3}
\end{equation}
Plugging the form of \cref{eq:solphi3} in \cref{eq:phi3eq}, we obtain for the particular solution
\begin{align}
    \phi_{31}^c(s)&=-\frac{\kappa B_0}{8}\left(\frac{5}{2}\frac{B_0'}{B_0}+\frac{k'}{k}+\frac{\eta'}{2}\cos 2 \delta-\mu u' \sin 2 \delta\right),\\
    \phi_{31}^s(s)&=\frac{\kappa B_0}{8}\left(-\tau+\frac{\eta'}{2}\sin 2 \delta+\mu u' \cos 2 \delta\right),
\end{align}
with both $\phi_{33}^c(s)$ and $\phi_{33}^s(s)$ integration constants from the homogeneous solution.

To obtain an expression for the solution of $\phi_n$ at arbitrary order in $\epsilon$, we first define the Laplacian operator $D$ in polar $(\rho,\omega)$ coordinates multiplied by $\rho^2$ as
\begin{equation}
    D \phi= \left[\rho \frac{\partial}{\partial \rho}\left(\rho \frac{\partial}{\partial \rho}\right)+\frac{\partial^2}{\partial \omega^2}\right]\phi=\sum_{n=2}^\infty\left(n^2 \phi_n+\ddot \phi_n\right)\epsilon^n \rho^n.
\end{equation}
Laplace's equation, \cref{eq:laplacephi}, can then be written as
\begin{equation}
    D \phi = -\sum_{n=2}^{\infty}\epsilon^n \rho^n\left[\frac{\kappa}{h_s}\left(\dot \phi_{n-1}\sin \theta-(n-1)\phi_{n-1}\cos \theta\right)+\frac{\phi_{n-2}''}{h_s^2}+\frac{(\kappa \cos \theta)'}{h_s^3}\phi_{n-3}'\right],
\end{equation}
where we defined $\phi_{-1}=0$.
We then expand the inverse powers of $h_s=1-\epsilon \kappa \rho \cos \theta$ in a power series in $\epsilon \rho$ and collect terms of the same power in $\epsilon \rho$, yielding for $n \ge 3$
\begin{align}
    n^2 \phi_n + \ddot \phi_n &= -\sum_{m=2}^n \kappa^{n-m} \cos \theta^{n-m}\left[\kappa\left(\dot \phi_{m-1} \sin \theta-(m-1)\phi_{m-1}\cos \theta\right)\right.\nonumber\\
    &\left.+(n-m+1)\phi_{m-2}''+\frac{(n-m+1)(n-m+2)}{2}(\kappa \cos \theta)' \phi_{m-3}'\right].
\label{eq:arbitraryorderphi}
\end{align}
We note that equation (\ref{eq:arbitraryorderphi}) has the form of a periodically driven harmonic oscillator with natural frequency $n$.
Similarly to $\phi_2$ and $\phi_3$, we decompose $\phi_n$ into its Fourier harmonics as
\begin{equation}
    \phi_n=\sum_{p=0}^n \phi_{np}^c(s) \cos pu+\phi_{np}^s(s)\sin p u.
\label{eq:phinp}
\end{equation}
At each order $n$, the expansion of \cref{eq:phinp} can be plugged in \cref{eq:arbitraryorderphi}, yielding a set of one dimensional differential equations for the coefficients  $\phi_{np}^c(s)$ and $\phi_{np}^s(s)$.
As the right-hand side of \cref{eq:arbitraryorderphi} only contains frequencies in $\omega$ up to $(n-2)$ (shown in Appendix \ref{app:Fourierphi}), the solutions of $\phi_{np}^c$ and $\phi_{np}^s$ in \cref{eq:phinp} for $0 \le p \le n-2$ are determined by the lower order $O(\epsilon^{n-1} \rho^{n-1})$ particular solutions.
The two remaining functions $\phi_{nn}^c(s)$ and $\phi_{nn}^s(s)$ are then integration parameters from the solution of the homogeneous equation in \cref{eq:phinp}.
Finally, we remark that the analicity condition of \cref{eq:analyticity} for $\phi$ can be derived from the solution of Laplace's equation, \cref{eq:laplacephi}, as shown in \cref{app:Fourierphi}.

\subsection{Vacuum Magnetic Toroidal Flux Surface Function}

We now determine the expression for the normalized vacuum toroidal flux $\psi^0(\rho,\omega,s)$.
Using \cref{eq:clebsch,eq:B0vac}, we determine $\psi^0$ via
\begin{equation}
    \nabla \phi \cdot \nabla \psi^0 = 0.
\label{eq:eqpsi0phi}
\end{equation}
Expanding the gradient operator using \cref{eq:gradientop}, we rewrite \cref{eq:eqpsi0phi} as
\begin{equation}
    \frac{\partial \phi}{\partial \rho}\frac{\partial \psi^0}{\partial \rho}+\frac{1}{\rho^2}\frac{\partial \phi}{\partial \omega}\frac{\partial \psi^0}{\partial \omega}+\frac{\epsilon^2}{h_s^2}\frac{\partial \phi}{\partial s}\frac{\partial \psi^0}{\partial s}=0.
\label{eq:eqpsi0phi1}
\end{equation}
An expression for the power series coefficients $\psi^0_{n}$ of $\psi^0$ can then be obtained by expanding both $\phi$ and $\psi^0$ in \cref{eq:eqpsi0phi1} in powers of $\epsilon \rho$ using \cref{eq:phirhoomega} and
\begin{equation}
    \psi^0=\sum_{n=0}^\infty \psi^0_n(\omega,s)\epsilon^n \rho^n,
\label{eq:psirhoomega}
\end{equation}
yielding
\begin{equation}
    2 n \phi_2 \psi^0_{n} + \dot \phi_2 \dot \psi^0_{n} + B_0 \psi^{0'}_{n}=F^0_{n},
\label{eq:psi0n}
\end{equation}
where $F^0_{0}=0$ and
\begin{align}
    F^0_{n}=-&\sum_{m=0}^{n-1}\left[(n+2-m)m\phi_{n+2-m}\psi^0_{m}+\dot \phi_{n+2-m}\dot \psi^0_{m}\right.\nonumber\\
    &\left.+\psi^{0'}_{m}\sum_{f=m}^{n}(n-f+1)\kappa^{n-f}\cos \theta^{n-f}\phi_{f-m}'\right],
\end{align}
for $n>0$.
Although a formal solution of \cref{eq:psi0n} can be obtained using the method of characteristics (as shown in \cref{app:characteristics}), here, we focus on deriving a one-dimensional system of differential equations for $\psi^0_n$ with the coefficients $\phi_{np}^c$ and $\phi_{np}^s$ as sources.
As an aside, we note that both $\psi^0$ and $\alpha^0$ obey the constraint in \cref{eq:eqpsi0phi}.
The distinction between the two is made by requiring $\psi^0$ to obey the analyticity condition, \cref{eq:analyticity}.

Using \cref{eq:psi0n} and the analyticity condition, \cref{eq:analyticity}, the lowest order solutions for $\psi^0$ are then given by $\psi^{0'}_0=\psi^0_1=0$.
The constant $\psi^0_0$ is set to zero by requiring $\psi^0$ to vanish on the magnetic axis.
Focusing on the $n=2$ case in \cref{eq:psi0n}, the equation for $\psi^0_{2}=\psi^0_{20} + \psi^{0s}_{22} \sin 2 u+\psi^{0c}_{22} \cos 2u$ can be written as
\begin{equation}
    \Psi^{0'}_{2}=A^0_2 \Psi^0_{2},
\label{eq:systempsi02}
\end{equation}
with $\Psi^0_{2}=B_0(s)^{-1}[\psi^0_{20} {,} \psi^{0s}_{22}{,} \psi^{0c}_{22}]^T$ and $A^0_{2}$ given by
\begin{equation}
    A^0_{2}= 
\left(
\begin{array}{ccc}
 0 & -2 \mu u' & \eta' \\
 -2 \mu u' & 0 & 2 u' \\
 \eta' & -2 u' & 0 \\
\end{array}
\right).
\end{equation}
The system of equations in \cref{eq:systempsi02} can be simplified by introducing the transformation
\begin{equation}
    \Psi^0_{2}=T_{2}\sigma^0_{2},
\label{eq:psitsig2}
\end{equation}
with $\sigma^0_{2}=[\sigma^0_{21} {,} \sigma^0_{22} {,}\sigma^0_{23}]^T$ and $T_{2}$ the matrix
\begin{equation}
    T_{2}= 2\left(
\begin{array}{ccc}
 - \sinh \eta & i &   \cosh \eta \\
  -\sinh \eta & - i & \cosh \eta \\
   \cosh \eta & 0 & -  \sinh \eta \\
\end{array}
\right).
\end{equation}
The quantities $\sigma^0_{2}$ then satisfy the following decoupled system of equations
\begin{equation}
    \sigma^{0'}_{2}=\left(
\begin{array}{cccc}
2\frac{i u'}{\cosh \eta} & 0 & 0 \\
0 & -2\frac{i u'}{\cosh \eta} & 0 \\
0 & 0 & 0
\end{array}
\right)\sigma^0_{2}.
\label{eq:sigma02}
\end{equation}
The solution of \cref{eq:sigma02} can then be given in terms of the integral
\begin{equation}
    v(s)=\int_0^s \frac{u'(x)}{\cosh \eta(x)}dx=\int_0^s \sqrt{1-\mu(x)^2}[\delta'(x)-\tau(x)]dx,
\label{eq:defvs}
\end{equation}
as
\begin{equation}
    \sigma^0_2(s)=[\sigma^{00}_{21} e^{2 i v(s)},\sigma^{00}_{22} e^{-2 i v(s)},\sigma^{00}_{23}]^T,
\label{eq:sigmasol2}
\end{equation}
with $\sigma^{00}_{21}, \sigma^{00}_{22}$ and $\sigma^{00}_{23}$ constants.
The flux surface function $\psi_2^0$ can then be found using \cref{eq:psitsig2}.

A more streamlined method to obtain the lowest order flux surface function $\psi^0_2$ can be found by noting that the free parameter $\delta$ in the analyticity condition in \cref{eq:analyticity} can be used to set $\psi^{0s}_{22}=0$, i.e., the expansion coefficients of the field $\phi$ are chosen in such a way that the  $\sin 2u$ terms in $\psi^0_2$ vanish.
From \cref{eq:systempsi02}, the remaining $\psi^{0}_{20}$ and $\psi^{0c}_{22}$ terms are then given by
\begin{align}
    \eta'&=\frac{\psi_{022}^{c'}}{\psi^0_{20}}=\frac{\psi^{0'}_{20}}{\psi^{0c}_{22}},\\
    \mu&=\tanh \eta=\frac{\psi^{0c}_{22}}{\psi^0_{20}}.
\end{align}
Defining $\psi^{0c}_{22}=\sinh \eta$ and $\psi^0_{20}=\cosh \eta$, the vacuum magnetic flux surface function $\psi^0{=\psi_2^0 \rho^2\epsilon^2+O(\epsilon^3)}$ can then be written as
\begin{equation}
    \psi_2^0=\frac{B_0 \pi}{\sqrt{1-\mu^2}}\left(1+\mu \cos 2 u\right){={B_0 \pi}\left(e^\eta \cos^2u+e^{-\eta} \sin^2 u\right)}.
\label{eq:psi02}
\end{equation}
The multiplicative constant in \cref{eq:psi02} is chosen such that $\psi$ equals the toroidal magnetic flux, i.e.,
\begin{equation}
    \psi = \frac{1}{L}\int (\mathbf B \cdot \nabla s) dV,
\label{eq:psinabla}
\end{equation}
with $dV$ the volume element, which in the $(\psi,\theta,s)$ coordinate system reads $dV=d\psi d\theta ds/(\nabla \psi \times \nabla \theta \cdot \nabla s)$.
We note that for a circular cross section with $\mu=0$, $\psi^0 = \pi B_0 \rho^2$.

The analysis that led to the solution in \cref{eq:sigmasol2} can be extended to higher orders.
{Explicit expressions for the third and fourth order solutions can be found in Appendix \cref{app:higherorder}.}
{Furthermore,} as shown in \cref{app:generalpsi}, the system of equations for $\Psi^0_n$, with $\Psi^0_n$ the column vector with the coefficients of the Fourier expansion of $B_0^{-n/2}\psi^0_n$ in $u$ for arbitrary $n$ with $\psi_n^0$ expanded as
\begin{equation}
    \psi_n^0=\sum_{p=0}^n \psi_{np}^{0c} \cos pu +\psi_{np}^{0s} \sin pu,
\end{equation}
can be cast into the following form
\begin{equation}
    \Psi_n^{0'}=A_n \Psi_n^0 + B_n^0,
\label{eq:odepsi0n}
\end{equation}
with $A_n$ and $B_n^0$ square matrices with periodic coefficients.
An analysis of the properties of $A_n$ and $B_n^0$ using the methods of Floquet theory is left for a future study.

The form of \cref{eq:sigma02,eq:sigma03eqs,eq:sigma04eqs} suggests the existence of a matrix $T_n$ at arbitrary order in $n$ such that the vectors $\sigma^0_n$ defined by
\begin{equation}
    \psi^0_n=T_n \sigma^0_n,
\label{eq:tonsigma}
\end{equation}
satisfy a decoupled system of first order differential equations.
Assuming the existence of $T_n$ for $n>4$, the decoupled system of equations for $\sigma^0_n=T_n^{-1}\psi^0_n$, a column vector with entries $\sigma^0_{nm}$ can be written for arbitrary $n$ in the following simplified form
\begin{equation}
    \sigma^{0'}_{nm}(s)-i m v'(s) \sigma^0_{nm} = F^0_{nm},
\label{eq:sigma0n}
\end{equation}
with $m$ an odd (even) integer if $n$ is odd (even) and $-n \le m \le n$ {and $v(s)$ given by \cref{eq:defvs}}.
The problem of determining the parameters associated with the flux surface function is then reduced to the solution of \cref{eq:sigma0n}.
A general solution of \cref{eq:sigma0n} is given by
\begin{equation}
    \sigma^0_{nm}(s)=e^{i m v(s)}\int_{s_0}^s F^0_{nm}(x)e^{-i m v(x)}dx.
\end{equation}
We now require $\psi^0$ (hence $\sigma^0_{nm}$) to be periodic on $s$ with period $L$, i.e., we impose the periodicity condition $\sigma^0_{nm}(s+L)=\sigma^0_{nm}(s)$, which sets the constant of integration $s_0$.
The periodic solution of \cref{eq:sigma0n} is then given by \citep{Mercier1964,Solovev1970}
\begin{equation}
    \sigma^0_{nm}(s)=\frac{e^{i m v(s)}}{e^{-i m v(L)}-1}\int_s^{s+L}e^{-i m v(x)}F^0_{nm}(x)dx.
\label{eq:solforsigma}
\end{equation}
Analysis of \cref{eq:solforsigma} shows that a periodic solution of $\sigma^0_{nm}$ (hence $\psi^0_n$) yields a resonant denominator when
\begin{equation}
    v(L)=2\pi\frac{l}{m},
\label{eq:resonancecond}
\end{equation}
with $l$ an integer.
We therefore conclude that for a solution of $\sigma_{nm}^0$ to exist, either the rotational transform on-axis is an irrational number, or the numerators in \cref{eq:solforsigma} vanish for a rational on-axis rotational transform.
In \cref{sec:rotationaltransform}, we relate the parameter $v(L)$ to the rotational transform $\iota$ and show that \cref{eq:resonancecond} is satisfied when the magnetic field lines in the vicinity of the magnetic axis close on themselves after one or more circuits along $s$, i.e., \cref{eq:resonancecond} is the condition for the existence of rational surfaces.

\subsection{Vacuum Field Line Label}
\label{sec:vacalpha}

The vacuum field line label $\alpha^0$ is found by equating \cref{eq:clebsch,eq:B0vac}, yielding the following set of three coupled equations
\begin{align}
    h_s \rho \frac{\partial \phi}{\partial \rho}&=\frac{\partial \psi^0}{\partial \omega}\frac{\partial \alpha^0}{\partial s}-\frac{\partial \psi^0}{\partial s}\frac{\partial \alpha^0}{\partial \omega},\label{eq:alphagen1}\\
    h_s \frac{\partial \phi}{\partial \omega}&=\rho \left( \frac{\partial \psi^0}{\partial s}\frac{\partial \alpha^0}{\partial \rho}-\frac{\partial \psi^0}{\partial \rho}\frac{\partial \alpha^0}{\partial s}\right),\label{eq:alphagen2}\\
    \frac{\epsilon^2\rho^2}{h_s} \frac{\partial \phi}{\partial s}&=\rho\left(\frac{\partial \psi^0}{\partial \rho}\frac{\partial \alpha^0}{\partial \omega}-\frac{\partial \psi^0}{\partial \omega}\frac{\partial \alpha^0}{\partial \rho}\right).\label{eq:alphagen3}
\end{align}
Expanding $\alpha^0$ in $\epsilon \rho$ as
\begin{equation}
    \alpha^0=\sum_n \alpha^0_n(\omega,s) \epsilon^n \rho^n,
\end{equation}
we find that, to lowest order in $\epsilon$, \cref{eq:alphagen1,eq:alphagen2,eq:alphagen3} reduce to
\begin{align}
    2\phi_2 &= \dot \psi^0_{2} \alpha^{0'}_0-\psi^{0'}_{2}\dot \alpha^0,\label{eq:alpha01}\\
    \dot \phi_2 &=- 2 \psi^0_{2} \alpha^{0'}_0,\label{eq:alpha02}\\
    \phi_0'&= 2 \psi^0_{2} \dot \alpha^{0}_0\label{eq:alpha03}.
\end{align}
We note that, by eliminating $\alpha^{0'}_0$ and $\dot \alpha^{0}_0$ in \cref{eq:alpha01} using \cref{eq:alpha02,eq:alpha03} we obtain \cref{eq:psi0n} with $n=2$.
Solving \cref{eq:alpha03} for $\alpha^{0}_0$ and plugging the result in \cref{eq:alpha02}, we find that
\begin{equation}
    \alpha^{0}=\frac{1}{2\pi}\left[\arctan \left(e^{-\eta} \tan u\right)-v(s)\right]+O(\epsilon).
\label{eq:alpha00}
\end{equation}
As expected, in contrast with $\psi^0$, $\alpha^0$ does not obey the analyticity condition in \cref{eq:analyticity}.

In order to obtain the vacuum field line label $\alpha^0$ to arbitrary order, we expand $\alpha^0$ in powers of $\epsilon \rho$ and obtain the following formulas for \cref{eq:alphagen2,eq:alphagen3} and $n>0$
\begin{align}
    -2 (\psi_2^0)^{n/2+1}\frac{\partial}{\partial s}\left(\alpha^0_{n} (\psi_2^0)^{-n/2}\right)&=\dot \phi_{n+2} - \kappa \cos \theta \dot \phi_{n+1}\nonumber\\
    &-\sum_{m=0}^{n-1}[m \alpha_m^0 \psi_{n+2-m}^{'0}-(n+2-m)\alpha_m^{'0} \psi_{n+2-m}^0],\label{eq:alphan2}\\
    2 (\psi_2^0)^{n/2+1}\frac{\partial}{\partial \omega}\left(\alpha^0_{n} (\psi_2^0)^{-n/2}\right)&=\sum_ {m=0}^n \phi_m' (\kappa \cos \theta) ^{n-m}\nonumber\\
    &+\sum_{m=0}^{n-1}[m \alpha_m^0 \dot\psi_{n+2-m}-(n+2-m)\dot \alpha_m^0 \psi_{n+2-m}^0].\label{eq:alphan3}
\end{align}
As done for $\alpha^0_0$, at each order, \cref{eq:alphan3} can be used to find an analytical expression for $\alpha^0_{n}$ up to an additive function of $s$, which is set by \cref{eq:alphan2}.
%

\section{MHD Equilibrium}
\label{sec:mhd}

We now solve \cref{eq:idealMHD} in the near-axis expansion formalism at first order in $\beta$.
In the following, the plasma current density $\mathbf J$ is normalized to $ \overline B/ 4 \pi R$ and $p$ to $\overline p$.
The linearized system of equations for $\mathbf B^1, \mathbf J$ and $\psi^1$ is given by the first order ideal MHD equation
\begin{equation}
    \mathbf J \times \mathbf B^0 =\beta p'(\psi^0) \nabla \psi^0,
\label{eq:idealmhd1}
\end{equation}
Ampère's law
\begin{equation}
    \nabla \times \mathbf B^1 = \mathbf J,
\label{eq:curlB}
\end{equation}
with both $\mathbf B^1$ and $\mathbf J$ divergence-free, i.e., $\nabla \cdot \mathbf B^1=\nabla \cdot \mathbf J=0$ and by the linearized flux surface condition
\begin{equation}
    \nabla \phi \cdot \nabla \psi^1 = -\mathbf B^1 \cdot \nabla \psi^0,
\label{eq:psi1}
\end{equation}
where we used the fact that $\mathbf B_0=\nabla \phi$.

We split the current density $\mathbf J$ into a parallel and perpendicular to $\mathbf B_0$ components
\begin{equation}
    \mathbf J =\beta p'(\psi^0)\left(\epsilon\mathbf J_\perp+J_\parallel \mathbf B^0\right).
\label{eq:j1}
\end{equation}
The multiplicative factor $\epsilon$ in \cref{eq:j1} is present in order to satisfy the divergence-free condition for the current density, $\nabla \cdot \mathbf J=0$.
The perpendicular current can be found using \cref{eq:idealmhd1}, yielding
\begin{equation}
    \mathbf J_\perp=\frac{\mathbf B^0 \times \nabla \psi^0}{|\mathbf B^0|^2},
\label{eq:jperpeq}
\end{equation}
while the parallel current is obtained by imposing $\nabla \cdot J=0$, yielding
\begin{equation}
    \mathbf B^0 \cdot \nabla J_\parallel = -\nabla \cdot \left(\frac{\mathbf B^0 \times \nabla \psi^0}{|\mathbf B^0|^2}\right)=\mathbf B^0 \cdot (\nabla B_0^{-2}\times \nabla \psi^0)=\mathbf J_\perp \cdot \frac{\nabla |\mathbf B^0|^2}{|\mathbf B^0|^2}.
\label{eq:parcurrent}
\end{equation}

\subsection{MHD Current Density Vector}

We start by deriving the forms of the perpendicular and parallel current densities, $\mathbf J_\perp$ and $J_\parallel$ respectively.
In the following, to simplify the notation, we define the coefficients of the inverse expansion for $|\mathbf B^0|^2$ as
\begin{equation}
    \frac{1}{|\mathbf B^0|^2}=\frac{1}{|\nabla \phi|^2}=\sum_{n=0}^\infty \frac{\mathcal{B}_n}{B_0(s)^2}\epsilon^n \rho^n,
\end{equation}
with $B_0(s)=|\mathbf B^0|_{\rho=0}$ the vacuum magnetic field modulus on axis and $\mathcal{B}_0$=1.

Using \cref{eq:jperpeq}, the three components $(J_{\perp s},J_{\perp \rho},J_{\perp \omega})$ of the perpendicular current can be written as
\begin{align}
    B_0(s)^2 J_{\perp s l+2}&=\sum_{g=0}^l \sum_{p=0}^g\mathcal{B}_{l-g}\left[(p+2)\phi_{p+2} \dot \psi_{0 g-p+2}-(g-p+2)\psi_{0 g-p+2} \dot \phi_{p+2}\right],\\
    B_0(s)^2 J_{\perp \rho l+1}&=\sum_{f=0}^l\sum_{g=0}^f \sum_{p=0}^g\mathcal{B}_{l-g}\left[\dot \phi_p \psi_{0 g-p+2}'-\phi_p' \dot \psi_{0 g-p+2}\right](\kappa \cos \theta)^{l-f},\\
    B_0(s)^2 J_{\perp \omega l+1}&=\sum_{f=0}^l\sum_{g=0}^f \sum_{p=0}^g\mathcal{B}_{l-g}\left[(g-p+2)\psi_{0 g-p+2}\phi_p' - p \psi_{0 g-p+2}' \phi_p\right](\kappa \cos \theta)^{l-f}.
\end{align}
We remark that, as expected, the perpendicular current vanishes on axis, i.e., $J_{\perp s 0}=J_{\perp \rho 0}=J_{\perp \omega 0}=0$.

The power series expansion coefficients $J_{\parallel n}$ of $J_\parallel=\sum_n J_{\parallel n} \epsilon^n \rho^n$ are obtained by expanding the terms included in \cref{eq:parcurrent} in power of $\epsilon \rho$, yielding
\begin{align}
    2n\phi_2 J_{\parallel n}+\dot \phi_2 \dot J_{\parallel n} +B_0 J_{\parallel n}'=\frac{G_n}{B_0(s)^2}-H_n,
\label{eq:jparf}
\end{align}
with $H_0=G_0=0$ and the coefficients $H_n$ and $G_n$ given by
\begin{equation}
    H_n=\sum_{g=0}^{n-1}\left[(n-g+2) \phi_{n-g+2} g J_{\parallel g}+\dot \phi_{n-g+2} \dot J_{\parallel g}+\phi'_{n-g}J_{\parallel g}' + (\kappa \cos \theta)^{n-g}\sum_{m=0}^g \phi'_{g-m}J_{\parallel m}'\right],
\end{equation}
and
\begin{align}
    G_n=&\sum_{g=1}^n \left\{\sum_{p=0}^{g-1}\sum_{r=0}^{n-g}\phi_r' (\kappa \cos \theta)^{n-g-r}\left[(g-p)\dot \psi_{0 p+2} \mathcal{B}_{g-p}-(p+2)\dot \psi_{0 p+2}\dot{\mathcal{B}}_{g-p}\right]\right.\nonumber\\
    &\left.+\sum_{l=1}^g\sum_{p=0}^{l-1} (\kappa \cos \theta)^{g-l}\left[\left(\mathcal{B}_{l-p-1}'-2\frac{B_0'}{B_0}\mathcal{B}_{l-p-1}\right)(p+2)\psi_{0 p+2}\dot \phi_{n-g+1}\right.\right.\nonumber\\
    &\left.\left.+\mathcal{B}_{l-p}\left(\psi_{0 p+1}' \phi_{n-g+1}-(l-p) \psi_{0 p+1} \dot \phi_{n-g+1}\right)\right]\right\},
\end{align}
respectively, for $n>0$.
We note that, as both $J_\parallel$ and $\psi^0$ are obtained by solving a magnetic differential equation, $J_{\parallel n}$ obeys an advection equation similar to the one of $\psi^0_{2}$, \cref{eq:psi0nadv}.
Therefore, similarly to \cref{eq:odepsi0n}, the components of the parallel current column vector $\mathcal{J}_{\parallel n}$ with $\mathcal{J}_{\parallel n}=[J_{\parallel n 0} J_{\parallel n 2}^c J_{\parallel n 2}^s ...]^T$ for even $n$ and $\mathcal{J}_{\parallel n}=[J_{\parallel n 1}^c J_{\parallel n 1}^s J_{\parallel n 3}^c J_{\parallel n 3}^s ...]^T$ for odd $n$, can be shown to obey
\begin{equation}
    \mathcal{J}_{\parallel n}'=A^0_{n}\mathcal{J}_{\parallel n}+C^0_{n},
\end{equation}
with $C^0_{n}$ a source term dependent on the components $G_n$ and $H_n$.
Using the same transformation matrix $T_{n}$ as in \cref{eq:tonsigma}, the components $\sigma^1_{nm}$ of $\sigma^1_{n}$, where $\sigma^1_{n}$ is given by
\begin{equation}
    \mathcal{J}_{\parallel n}=T_{n} \sigma^1_{n},
\end{equation}
can be shown to satisfy \cref{eq:sigma0n} with a different source term, namely
\begin{equation}
    \sigma^{1'}_{nm}(s)-i m v' \sigma^1_{nm} = D^1_{nm},
\label{eq:sigma1n}
\end{equation}
with $D^1_{nm}=T_{n}^{-1}C^0_{n}$.
The solution of $\sigma^1_{nm}$ is then of the form of \cref{eq:solforsigma}.

For $f=0$, \cref{eq:jparf} determines the current on axis $J_{\parallel 0}$ to be a constant.
For $f=1$, the equation for the coefficients $\lambda$ and $\mu$ of the first order current $J_{\parallel 1}=\lambda \sin u + \mu \cos u$ can be written as
\begin{equation}
    \sigma_1'-i v'(s) \sigma_1 = -B_0^{3/4}(s) 4 \kappa \left[e^{\eta/2} (1-\mu \cos \delta) + i e^{-\eta/2}(1+\mu)\sin \delta\right],
\end{equation}
with $\sigma_1 = B_0(s)^{-1/2}(\lambda e^{\eta/2}-i \mu e^{-\eta/2})$.

\subsection{MHD Magnetic Field Vector}

We now proceed by calculating the first order magnetic field by expanding the components of $\mathbf B^1$ in powers of $\epsilon \rho$ as
\begin{equation}
    \mathbf B^1=\sum_n \mathbf B^1_n(s \omega) \epsilon^n \rho^n
\end{equation}
and use \cref{eq:curlB} to solve for $\mathbf B^1_{\rho n}, \mathbf B^1_{\omega n}$ and $\mathbf B^1_{sn}$.
From the $\rho$ component of \cref{eq:curlB} we find at lowest order the constraint constraint $\dot B^1_{s0}=0$.
For higher order, we obtain for the $\rho$ component
\begin{equation}
\sum_{n=1}^\infty \epsilon^n \rho^n \left[\frac{\partial B^1_{sn}}{\partial \omega}-\kappa \frac{\partial}{\partial \omega}(\cos \theta B^1_{sn-1})-\frac{\partial B^1_{\omega  n-1}}{\partial s}\right]=A^J_\rho,
\label{eq:curlbrho}
\end{equation}
with $A^j_\rho=\beta h_s p'(\psi^0) (\epsilon \rho J_{\perp \rho} + J_\parallel \rho \partial \phi/\partial \rho)$, for the  $\omega$ component
\begin{equation}
    \sum_{n=1}^{\infty}\epsilon^n \rho^n\left[n B^1_{sn}-(n+1)\kappa \cos \theta B^1_{sn-1} - \frac{\partial B^1_{\rho n-1}}{\partial s}\right]=A^J_{\omega}
\label{eq:curlbw}
\end{equation}
with $A^J_\omega=- \beta h_s p'(\psi^0)(\epsilon \rho J_{\perp \rho}+J_\parallel {\partial \phi}/{\partial \omega})$ and for the $s$ component
\begin{equation}
    \sum_{n=0}^\infty \epsilon^n \rho^n \left[(n+1)B^1_{\omega n}-\frac{\partial B^1_{\rho n}}{\partial \omega}\right]=A^J_s,
\label{eq:curlbs}
\end{equation}
with $A^J_s=\epsilon \rho \beta p'(\psi^0)(J_{\perp s}+ h_s^{-1} J_\parallel \partial\phi/\partial s)$.
In order to simplify the calculations and eliminate one of the three components of \cref{eq:curlB}, we replace \cref{eq:curlbrho} by the condition, $\nabla \cdot \mathbf B^1 = 0$, which can be written as
\begin{equation}
    \frac{\partial (\rho B^1_{\rho})}{\partial \rho}+\frac{\partial B^1_{\omega}}{\partial \omega}=-\epsilon \rho \left[\frac{\partial B^1_{s}}{\partial s}+\kappa \frac{\partial (\cos \theta B^1_{\omega})}{\partial \omega}+\frac{\kappa \cos \theta}{\rho}\frac{\partial (\rho^2 B^1_{s})}{\partial \rho}\right].
\label{eq:divbrhows}
\end{equation}
Expanding the functions $A^J_\omega$ and $A^J_s$ in terms of powers of $\epsilon \rho$, the following expressions for the perturbed magnetic field are found
\begin{align}
    (n+1)^2 B^1_{\rho n}+\frac{\partial^2 B^1_{\rho n}}{\partial \omega^2}&=-(n+1)\left[\frac{1}{n+1}\frac{\partial A^J_{sn}}{\partial \omega}+\frac{\partial B^1_{sn-1}}{\partial s}\right.\nonumber\\
    &\left.+\kappa\frac{\partial (\cos \theta B^1_{\omega n-1})}{\partial \omega}+(n+2)\kappa \cos \theta B^1_{sn-1}\right],\label{eq:brho1eq}\\
    B^1_{\omega n}&=\frac{1}{n+1}\left(A^J_{sn}+\frac{\partial B_{1 \rho n}}{\partial \omega}\right),\label{eq:bomega1eq}\\
    B^1_{sn}&=\frac{1}{n}\left[\frac{\partial B^1_{\rho n-1}}{\partial s}+(n+1)\kappa \cos \theta B^1_{sn-1} + A^J_{\omega n}\right].\label{eq:bs1eq}
\end{align}
The lowest order field $\mathbf B^1_{0}$ is found to satisfy $B^1_{\rho 0}=B^{1c}_{\rho 0} (s) \cos u + B^{1s}_{\rho 0} (s) \sin u$, $B^1_{\omega 0}=\partial B^1_{\rho 0}/\partial \omega$ and $B^1_{s0}=B^1_{s0}(s)$, in agreement with \citet{Solovev1970}.
While analyticity could be proven rigorously for the vacuum case, we note that the right-hand side of the forced linear harmonic oscillator equation for $B_{\rho n}^1$ in \cref{eq:brho1eq} might contain $n+1$ resonating frequencies resulting from the product $J_\parallel \partial \phi/\partial s$ in the $A_s^J$ term.
These resonances can lead to the appearance of non-analytic and weakly singular terms of the form $\rho^n(\log{\rho})^m$ as discussed in \cite{Weitzner2016}.

\subsection{MHD Flux Surface Function}

We now obtain $\psi^1$ using the linearized flux surface condition, \cref{eq:psi1}.
We take advantage of the fact that \cref{eq:psi1} is similar to \cref{eq:eqpsi0phi}, although with a non-zero source term, i.e.,
\begin{equation}
    2n \phi_2 \psi^1_{n} + \dot \phi_2 \dot \psi^1_{n}+B_0 \psi^1_{n} = F^1_{n},
\label{eq:psin1}
\end{equation}
with
\begin{equation}
    F^1_{n}=F^0_{n}- \frac{\epsilon}{n!}\left[\frac{1}{\epsilon^2}\frac{\partial^n}{\partial \rho^n}\left(B^1_{\rho} \frac{\partial \psi^{0}}{\partial \rho}+\frac{B^1_{\omega}}{\rho}\frac{\partial \psi^0}{\partial \omega}+\epsilon \frac{B^1_{s}}{h_s} \frac{\partial \psi^0}{\partial s}\right)\right]_{\rho=0}.
\end{equation}
By plugging the analyticity condition for $\psi^1$, \cref{eq:analyticity}, in \cref{eq:psin1} and defining the column vector $\Psi_n^1$ in an analogous manner with $\Psi_n^0$, the following set of equations for $\psi^1_{n}$ is found
\begin{equation}
    \Psi_n^{1'}=A_n\Psi_n^{1}+B_n^1,
\end{equation}
with $B_n^1$ the matrix of Fourier coefficients of $F^1_n$ at each order $n$.
The components of $A_n$ are given in \cref{app:generalpsi}.

The lowest order solution for $\psi^{1}$ is given by
\begin{equation}
    \psi^1=\epsilon \rho [\psi_{11}^c(s) \cos u + \psi_{11}^c(s) \sin u] + O(\epsilon^2)
\label{eq:psi1mhd}
\end{equation}
with $\psi_{11}^c$ and $\psi_{11}^s$ obeying the set of equations
\begin{align}
    \frac{B_0(s)}{2}\psi_{11}^{c'}=\phi_{22}^s \psi_{11}^s+(\phi_{20}+\phi_{22}^c)\psi_{11}^c-B_{1\rho}^c(\psi^0_{20} + \psi^{0c}_{22})-B_{1\rho}^s \psi^{0s}_{22},\\
    \frac{B_0(s)}{2}\psi_{11}^{s'}=\phi_{22}^s \psi_{11}^c+(\phi_{20}-\phi_{22}^c)\psi_{11}^s-B_{1\rho}^s(\psi^0_{20} - \psi^{0c}_{22})-B_{1\rho}^c \psi^{0s}_{22}.
\end{align}

\subsection{Shafranov Shift}

Here, we show how to obtain the position of the magnetic axis $(x_M, y_M)$ once finite $\beta$ effects are taken into account.
Although the procedure is valid for arbitrary order in the expansion parameter $\epsilon$, we calculate $(x_M, y_M)$ explicitly to lowest order in $\epsilon$ only.
We first rewrite $\psi^1$ in Cartesian $(x,y)$ coordinates as
\begin{equation}
    \psi^1=\epsilon(a x + b y),
\end{equation}
with the functions $a$ and $b$ given by $a=\psi_{11}^c \cos \delta + \psi_{11}^s \sin \delta$ and $b=-\psi_{11}^c \sin \delta + \psi_{11}^s \cos \delta$.
The lowest order magnetic flux surface function in $\epsilon$ and $\beta$, in $x$ and $y$ coordinates, is then given by
\begin{equation}
    \psi=\psi^0+\psi^1=a x + b y + A x^2 + B y^2 + C x y.
\end{equation}
with $A=(1+\mu \cos 2 \delta)/\sqrt{1-\mu^2}$, $B=(1-\mu \cos 2 \delta)/\sqrt{1-\mu^2}$ and $C=-2 \mu \sin 2 \delta/\sqrt{1-\mu^2}$
Setting the derivatives of $\psi$ with respect to $x$ and $y$ equal to zero, the position of the magnetic axis $(x_M,y_M)$ is found to be
\begin{equation}
    x_M=\frac{2aB-bC}{C^2-4AB},
\end{equation}
and
\begin{equation}
    y_M=\frac{2A b - a C}{C^2+4AB},
\end{equation}
The condition that the distortion of the magnetic surfaces be small, i.e., $x_M \sim y_M \ll a$ leads to a limit on the maximum allowed $\beta$.
For the derivation of the plasma $\beta$-limit $\beta \ll 2 \epsilon \iota_0^2$ for the particular case of a circular magnetic axis, $\delta=n \pi$ and neglecting curvature effects, see \citet{Solovev1970}.

\subsection{MHD Field Line Label}

Finally, for the field line label $\alpha^1$, using $\mathbf B^1= \nabla \psi^1 \times \nabla \alpha^1$, we derive the following set of three coupled equations
\begin{align}
    h_s  \epsilon \rho B_{1\rho}&=\frac{\partial \psi^1}{\partial \omega}\frac{\partial \alpha^1}{\partial s}-\frac{\partial \psi^1}{\partial s}\frac{\partial \alpha^1}{\partial \omega},\label{eq:alphagen11}\\
    h_s \epsilon \rho B^1_{\omega}&=\rho \left( \frac{\partial \psi^1}{\partial s}\frac{\partial \alpha^1}{\partial \rho}-\frac{\partial \psi^1}{\partial \rho}\frac{\partial \alpha^1}{\partial s}\right),\label{eq:alphagen12}\\
    \epsilon^2\rho^2 B^1_{s}&=\rho\left(\frac{\partial \psi^1}{\partial \rho}\frac{\partial \alpha^1}{\partial \omega}-\frac{\partial \psi^1}{\partial \omega}\frac{\partial \alpha^1}{\partial \rho}\right).\label{eq:alphagen13}
\end{align}
Expanding $\alpha^1$ in a series of powers of $\epsilon \rho$ and following the approach in \cref{sec:vacalpha}, we find
\begin{align}
    -(\psi_{1}^1)^{n+1}\frac{\partial}{\partial s}\left[\alpha^1_{n} (\psi_{1}^1)^{-n}\right]&=B^1_{\omega n} - k \cos \theta B^1_{\omega n-1}\nonumber\\
    &-\sum_{m=0}^{n-1}\left[\psi_{n-m+1}^{1'}m\alpha^1_{m}-\psi^1_{ n-m+1}(n-m+1)\alpha^{1'}_{m}\right]\label{eq:alphan12}\\
    (\psi_{1}^1)^{n+1}\frac{\partial}{\partial \omega}\left[\alpha^1_{n} (\psi_{1}^1)^{-n}\right]&=B_{1s n-1}\nonumber\\
    &+\sum_{m=0}^{n-1}\left[\dot \psi^1_{n-m+1}m\alpha^1_{m}-\psi^1_{ n-m+1}(n-m+1)\dot \alpha^1_{m}\right].\label{eq:alphan13}
\end{align}
Analogously to \cref{eq:alphan2,eq:alphan3}, at each order in $\epsilon^n$, \cref{eq:alphan13} can be used to find an analytical expression for $\alpha^1_{n}$ up to an additive function of $s$, which is set by \cref{eq:alphan3}.

\section{Rotational Transform}
\label{sec:rotationaltransform}

In this section, we aim at calculating the rotational transform $\iota$ at arbitrary order, given an {explicit form} of $\alpha(\rho,\omega,s)$.
We first note that, as $\mathbf B = \nabla \psi \times \nabla \alpha$ must be periodic in both $s$ and $\theta$, the most general form for the field line label $\alpha$ is given by
\begin{equation}
    \alpha = f(\psi) \theta - g(\psi) s + \tilde \alpha(\psi, \theta, s),
\label{eq:alphagf}
\end{equation}
with $\tilde \alpha$ a periodic function in both $\psi$ and $\theta$.
The functions $g(\psi)$ and $f(\psi)$ can be found using the expression for the toroidal flux in \cref{eq:psinabla} and the specific poloidal flux
\begin{equation}
    \chi'(\psi) = \frac{d}{d\psi}\left[\frac{1}{2\pi}\int (\mathbf B \cdot \nabla \theta) dV\right] = {\iota-N},
\label{eq:chinabla}
\end{equation}
with $\chi=\chi(\psi)$ \citep{Kruskal1958,d1991flux} {and $N$ the total number of rotations of the normal vector after one circuit along the axis}.
Here, the poloidal magnetic flux $\chi$ is given by the flux that passes through the two closed curves given by the magnetic axis and the line 
\begin{equation}
    \theta = \omega - \gamma(s) = \text{const},~\psi(\rho,\theta,s)=\text{const},
\label{eq:secondcurve}
\end{equation}
i.e., the trace of the intersection of the magnetic surface normal to the magnetic axis.
We note that in order to calculate the angle through which the magnetic field line rotates around the axis for a complete circuit along the torus, we subtracted from $\chi'(\psi)$ the number of times $N$ the curve in \cref{eq:secondcurve} (or equivalently the normal vector $\mathbf n$) encircles the magnetic axis.
Using \cref{eq:alphagf}, we can write the magnetic field $\mathbf B$ as
\begin{equation}
    \mathbf B = f(\psi) \nabla \psi \times \nabla \theta - g(\psi) \nabla \psi \times \nabla s +\nabla \psi \times \nabla \tilde \alpha.
\label{eq:bnablanabla}
\end{equation}
Plugging \cref{eq:bnablanabla} in \cref{eq:psinabla}, we find $f(\psi)=1/2\pi$, while using \cref{eq:bnablanabla,eq:chinabla} we find $g(\psi)=(\iota-N)/L$, yielding
\begin{equation}
    \alpha=\frac{\theta}{2\pi}-(\iota-N)\frac{s}{L}+\tilde \alpha,
\end{equation}
{with $L$ defined in \cref{eq:unitangent} as the total length of the axis.}
An expression for $\iota$ in terms of $\alpha$, valid at arbitrary order in $\epsilon$, is then given by
\begin{equation}
    \iota={\alpha(s,\theta)-\alpha(s+L,\theta)}+N.
\label{eq:iotaarbitraryorder}
\end{equation}

We now apply \cref{eq:iotaarbitraryorder} to the lowest order expression of $\alpha$ in \cref{eq:alpha00}, yielding
\begin{align}
    \iota &= \frac{1}{2\pi}\left(-\left.\arctan \left[e^{-\eta(s')} \tan(\theta + \delta(s')) \right]\right._{s'=s}^{s'=s+L}+\left[v(s+L)-v(s)\right]\right)+N\nonumber\\
    &=\frac{1}{2\pi}\left(v(L)-\left[\delta(L)-\delta(0)\right]\right)+N=\frac{1}{2\pi}\left(\int_{0}^L\left[(\delta'-\tau)\sqrt{1-\mu^2}\right] ds-\left[\delta(L)-\delta(0)\right]\right)+N. \label{eq:iota0}
\end{align}
Note that in the case $\delta(L)=\delta(0)$ and $N=0$, the resonance condition in \cref{eq:resonancecond} is equivalent to the condition of $\iota=l/m$ with $l$ and $m$ integers.
In order to interpret the rotational transform obtained in \cref{eq:iota0}, we write the lowest order toroidal flux $\psi^0_{2}$ as
\begin{equation}
    \frac{\psi^0_{2}}{B_0 \pi}=X^2+Y^2 = Ax^2 + By^2 +C xy,
\label{eq:psixy}
\end{equation}
where $(X,Y)$ are the elliptical coordinates
\begin{equation}
    X=e^{\eta/2} \rho \cos u,~Y=e^{-\eta/2} \rho \sin u,
\end{equation}
$(x,y)$ Cartesian coordinates in the plane locally perpendicular to the magnetic axis given by \cref{eq:poscart}.
From \cref{eq:psixy}, it is clear that surfaces of constant $\psi^0_{2}$ are circles in elliptical coordinates and ellipses in Cartesian coordinates.
We then conclude that the total turning angle of a field line after one toroidal rotation, i.e., the rotational transform, is given by the sum of the total turning angle $\arctan Y/X^{s=L}_{s=0}$ at constant field line label $\alpha^{0}_0$, with the total rotation angle of the ellipse $\zeta$ in the $(x,y)$ plane and with the total rotation angle of the $(x,y)$ plane itself, i.e., the number of times $N$ the curve in \cref{eq:secondcurve} encircles the magnetic axis.
As $\alpha^{0}_0$ can be written as
\begin{equation}
    \alpha^{0}_0=\arctan \frac{Y}{X}-v(s),
\end{equation}
the total turning angle $\arctan Y/X|^{s=L}_{s=0}$ at constant $\alpha_0^0$ is then given by $v(L)-v(0)=v(L)$.
The rotating angle $\zeta$ of the ellipse can be determined via the relations $A=\cos^2 \zeta/a^2+\sin^2 \zeta/b^2$, $B=\sin^2 \zeta/a^2+\cos^2 \zeta/b^2$ and $C=\sin 2 \zeta/a^2-\sin 2 \zeta/b^2$, which applied to \cref{eq:psixy} yields $\zeta=-\delta$.
The rotational transform is then given by the ratio between the total summing angles and the angle $2\pi$ related to one complete toroidal revolution, yielding
\begin{equation}
    \iota=\frac{v(L)-\left[\delta(L)-\delta(0)\right]}{2\pi}+N,
\end{equation}
in agreement with the result in \cref{eq:iota0}.

\section{Comparison With the Garren-Boozer Construction}
\label{sec:comparisongb}

In this section, we show the equivalence between the near-axis framework developed in the previous sections using the direct method, with the Garren-Boozer construction based on the inverse coordinate approach \citep{Garren1991}.
For simplicity, we perform the comparison between explicit vacuum solutions of the direct and inverse approach up to first order in $O(\epsilon)$.
In Boozer coordinates $(\psi,\vartheta,\varphi)$, the magnetic field can be written in a contravariant representation as
\begin{equation}
    \mathbf B = (\nabla \psi \times \nabla \vartheta + \iota(\psi)\nabla \varphi \times \nabla \psi)/2\pi,
\label{eq:boozerBcon}
\end{equation}
while in a covariant representation it reads
\begin{equation}
    \mathbf B = \overline \beta(\psi,\vartheta,\varphi) \nabla\psi/2\pi + I(\psi) \nabla \vartheta + G(\psi) \nabla \varphi,
\label{eq:boozerBcov}
\end{equation}
where $\overline \beta$ is related to the Pfirsch-Schl\" uter current \citep{Boozer1981}, $I(\psi)$ is $\mu_0/(2\pi)$ times the toroidal current enclosed by the flux surface and $G(\psi)$ is $\mu_0/(2\pi)$ times the poloidal current outside the flux surface.
Focusing on the vacuum case, the covariant representation in \cref{eq:boozerBcov} can be written as
\begin{equation}
    \mathbf B = G_0 \nabla \varphi,
\label{eq:boozerBcovvac}
\end{equation}
with $G_0$ a nonzero constant given by
\begin{equation}
    G_0=\frac{L}{\int_0^{2\pi}d\varphi B_0^{-1}(\varphi)},
\end{equation}
or, alternatively, $s'(\varphi)=G_0/B_0$ with $s$ the arclength function.
The Jacobian $\sqrt{g}$ can be found from the product of \cref{eq:boozerBcon,eq:boozerBcovvac}, yielding
\begin{equation}
    \sqrt{g}=\frac{1}{\nabla \psi \cdot \nabla \vartheta \times \nabla \varphi}=\frac{G_0}{2 \pi B^2}.
\end{equation}
We note that the normalization constant $\overline B$ in \citet{Landreman2018} corresponds to the constant $\overline B$ used here to normalize $\psi$ and $\bf B$ times $1/\pi$ due to the different definitions of $\psi$.

The direct transformation from $(\psi,\vartheta,\varphi)$ to $(\rho,\omega,s)$ coordinates can be found in the following way.
From the equality between \cref{eq:B0vac,eq:boozerBcovvac}, the toroidal Boozer angle can be computed at any order in vacuum using
\begin{equation}
    \varphi(\rho,\omega,s)=\frac{\phi}{G_0}.
\label{eq:varphieq}
\end{equation}
The toroidal flux $\psi(\rho,\omega,s)$ is computed using \cref{eq:psi02}, while the poloidal Boozer angle $\vartheta$ can be found by first noting that the magnetic field $\mathbf B$ in \cref{eq:boozerBcon} can be written as $\mathbf B=\nabla \psi \times \nabla(\vartheta - \iota \varphi)/2\pi=\nabla \mathbf \psi \times \nabla \alpha$ yielding
\begin{equation}
    \vartheta=2\pi \alpha+\iota \varphi.
\label{eq:boozerthetaeq}
\end{equation} 
and plugging the expressions for $\alpha$ and $\varphi$ from \cref{eq:varphieq,eq:alpha00} in \cref{eq:boozerthetaeq}.
The transformation between both coordinate systems is then given by
\begin{align}
    \psi(\rho,\omega,s) &= \frac{B_0 \pi \rho^2}{\sqrt{1-\mu^2}}[1+\mu(s)\cos (2u)]={B_0 \pi \rho^2}\left(e^{\eta} \cos^2 u+e^{-\eta}\sin^2 u\right)+O(\rho^3)\label{eq:psiboozer}\\
    \varphi(\rho,\omega,s) &= \frac{1}{G_0}\int_0^s B_0(s') ds'+O(\rho),\label{eq:phiboozer}\\
    \vartheta(\rho,\omega,s) &= \arctan\left(e^{-\eta}\tan u\right)- v(s)+\frac{\iota}{G_0}\int_0^s B_0(s') ds'+O(\rho).\label{eq:thetaboozer}
\end{align}

We now show the equivalence of the first-order position vector $\mathbf r$ between the direct and inverse approaches.
This is done first by stating the solution for $\mathbf r$ in the Garren-Boozer construction and its related constraints, and showing that the lowest order transformation in \cref{eq:psiboozer,eq:phiboozer,eq:thetaboozer} together with the results from the previous sections yields a similar set of constraints.
In the Garren-Boozer construction, to first order in $\epsilon, $ the position vector is given by
\begin{equation}
    \mathbf r = \mathbf r_0(s)+X_1 \mathbf n(s)+Y_1 \mathbf b(s),
\label{eq:posrboozer}
\end{equation}
with
\begin{equation}
    X_1=\sqrt{\psi}\left[X_{1,1c}(\varphi) \cos \vartheta + X_{1,1s}(\varphi) \sin \vartheta\right],
\end{equation}
and
\begin{equation}
    Y_1=\sqrt{\psi}\left[Y_{1,1c}(\varphi) \cos \vartheta + Y_{1,1s}(\varphi) \sin \vartheta\right].
\end{equation}
The first constraint is given by
\begin{equation}
    X_{1,1c}Y_{1,1s}-X_{1,1s}Y_{1,1c}=\frac{1}{\pi B_0}.
\label{eq:constraintgb1}
\end{equation}
The second constraint is the solution for the magnetic field strength $B=B_0(1-\kappa X_1)$.
Finally, the constraint equation derived from the $\mathbf n$ and $\mathbf b$ equations at $O(\epsilon)$, i.e., Eq. (63) of \citet{Garren1991a} and Eq. (3.8) of \citet{Landreman2018}, reads
\begin{equation}
    \iota V_1= X_{1,1c}\frac{d X_{1,1s}}{d\varphi}-X_{1,1s}\frac{d X_{1,1c}}{d\varphi}+Y_{1,1c}\frac{d Y_{1,1s}}{d\varphi}-Y_{1,1s}\frac{d Y_{1,1c}}{d\varphi}-\frac{2 G_0}{\pi B_0^2} \tau,
\label{eq:gbconstraint1}
\end{equation}
with
\begin{equation}
    V_1=X_{1,1s}^2+X_{1,1c}^2+Y_{1,1c}^2+Y_{1,1s}^2.
\end{equation}
In the following, we show the equivalence of the three constraints between the direct and inverse approaches.

We equate \cref{eq:posMercier,eq:posrboozer} and express the Boozer angle $\vartheta$ in terms of $(\rho,\omega,s)$ using \cref{eq:thetaboozer}, yielding the following expressions for $(X_{1,1c},Y_{1,1s},Y_{1,1c},Y_{1,1s})$
\begin{align}
    X_{1,1c}&=\frac{1}{\sqrt{B_0 \pi}}\left(e^{-\eta/2}\cos \delta \cos f - e^{\eta/2} \sin \delta \sin f\right),\label{eq:x11c}\\
    X_{1,1s}&=\frac{1}{\sqrt{B_0 \pi}}\left(e^{-\eta/2}\cos \delta \sin f + e^{\eta/2} \sin \delta \cos f\right),\label{eq:x11s}\\
    Y_{1,1c}&=\frac{-1}{\sqrt{B_0 \pi}}\left(e^{\eta/2}\cos \delta \sin f + e^{-\eta/2} \sin \delta \cos f\right),\label{eq:y11c}\\
    Y_{1,1s}&=\frac{1}{\sqrt{B_0 \pi}}\left(e^{\eta/2}\cos \delta \cos f - e^{-\eta/2} \sin \delta \sin f\right)\label{eq:y11s},
\end{align}
where $f(s)=\iota\varphi(s)-v(s)$.
In order to derive \cref{eq:x11c,eq:x11s,eq:y11c,eq:y11s}, we have expressed the toroidal flux as $\psi=B_0 \pi \rho^2(e^\eta \cos^2 u + e^{-\eta} \sin^2 u)$ and used the trigonometric identities
\begin{equation}
    \cos\left[\arctan(e^{-\eta} \tan u)\right]=\frac{e^{\eta/2}\cos u}{\sqrt{e^{\eta}\cos^2 u + e^{-\eta}\sin^2 u}},
\end{equation}
and
\begin{equation}
    \sin\left[\arctan(e^{-\eta} \tan u)\right]=\frac{e^{-\eta/2}\sin u}{\sqrt{e^{\eta}\cos^2 u + e^{-\eta}\sin^2 u}}.
\end{equation}
Plugging \cref{eq:x11c,eq:x11s,eq:y11c,eq:y11s} in \cref{eq:constraintgb1}, the first Garren-Boozer constraint is automatically satisfied.
The second constraint related to the magnetic field modulus is  satisfied since $X_1=\rho \cos \theta$ and the lowest order vacuum magnetic field in the direct approach from $B=|\nabla \phi|$ is given by $B=B_0(1-\kappa \rho \cos \theta)$.
Finally, using the system of \cref{eq:x11c,eq:x11s,eq:y11c,eq:y11s}, the constraint in \cref{eq:gbconstraint1} is also satisfied automatically.

\section{Numerical Comparison with W7-X Equilibrium}
\label{sec:numerical}

With the aim of testing the framework developed in the previous sections for a realistic equilibrium, we now focus on describing the inner surfaces of the optimized stellarator W7-X using the near-axis expansion, and evaluate the accuracy of the expansion as we move radially outward towards increasing $\rho$.
{
For this study, the vacuum W7-X standard configuration is used, which corresponds to the A configuration of \citet{Geiger2015} at $\beta=0$.}
{As a boundary}, we choose a W7-X surface with a magnetic toroidal flux (in SI units) of $\psi = 0.01$ T m$^2$.
We remark that the toroidal flux in the plasma boundary for this configuration is $\psi_b = 2.19$ T m$^2$, yielding $\psi/\psi_b \simeq 4.6 \times 10^{-3}$ for the surface considered here.
The expansion parameter on this particular surface can be estimated as $\epsilon \rho \sim \sqrt{\psi/\overline B R^2} \sim 10^{-2}$, with $\overline B \sim 3$ T and $R \sim 5.5$ m.
For simplicity, we use the lowest order expressions in vacuum for $\psi$, i.e., \cref{eq:psi02,eq:psi03}, and perform a nonlinear least-squares fit to find the functions $\mu, \delta, B_0, \psi_{31}^{0c}, \psi_{31}^{0s}, \psi_{33}^{0c}$ and $\psi_{33}^{0s}$ that best approximate the shape of the magnetic field near the axis of W7-X.
{The numerical tool used for this study can be found in \citet{SENAC}.}
As inputs for the numerical procedure, we use the magnetic axis of W7-X and the Fourier harmonics associated with that particular surface of constant $\psi$, as given by the VMEC code \citep{Hirshman1983}.
In VMEC, a cylindrical coordinate system is employed, in which the  position vector $\mathbf r$ is written as
\begin{equation}
    \mathbf r = R\mathbf e_R(\Phi)+Z\mathbf e_Z,
\label{eq:rposcyl}
\end{equation}
with $(\mathbf e_R, \mathbf e_\Phi, \mathbf e_Z)$ the cylindrical unit basis vectors and $(R,\Phi,Z)$ standard cylindrical coordinates.
The two coordinates used to parametrize the flux surface in VMEC are a poloidal angle $\theta_V$ and the standard toroidal angle $\Phi$.
Assuming stellarator geometry, the radial and vertical components of $\mathbf r$ can then be written as
\begin{equation}
    R=\sum_{m,n}R_{mn}\cos(m \theta_V - n \Phi),
\end{equation}
and
\begin{equation}
    Z=\sum_{m,n}Z_{mn}\sin(m \theta_V - n \Phi).
\end{equation}
The magnetic axis is also described using a cylindrical coordinate system, with $R, Z$ and $\mathbf e_R$ parametrized using a single quantity $\Phi_a$, satisfying $0 \le \Phi_a < 2\pi$.
{The magnetic axis $\mathbf r_0(\phi)$ of the W7-X configuration used here is given by
\begin{equation}
\begin{split}
	\mathbf r_0(\Phi_a) &= \left[5.56+0.37 \cos(5\Phi_a)+0.02 \cos(10\Phi_a)\right]\mathbf e_R\\
	&-\left[0.31 \sin(5\Phi_a)+0.02 \sin(10\Phi_a)\right] \mathbf e_Z.
\end{split}
\end{equation}
}
%
%
%

We start by deriving the relation between Mercier's coordinates $(\rho, \theta, s)$ and VMEC's poloidal $\theta_V$ and toroidal $\Phi$ coordinates.
This allows us to parametrize the surfaces of constant $\psi$ in terms of $\rho(\psi, \theta_V, \Phi)$ and find a parametric form for the position vector $\mathbf r$ in \cref{eq:posMercier} in terms of $(\theta_V, \Phi)$ at any order in $\epsilon$.
Starting with the arclength function $s$, we use the fact that the tangent vector $\mathbf t$ is a unit vector and employ the chain rule in \cref{eq:unitangent}, yielding $ds/d\Phi_a=|{d \mathbf r_0}/{d \Phi_a}|$.
Next, the relation between the toroidal angle on axis $\Phi_a$ and the poloidal and toroidal angles on the surface $(\theta_V, \Phi)$ is found by imposing that the tangential component of $\mathbf r - \mathbf r_0$ vanishes, i.e.,
\begin{equation}
    \mathbf t(\Phi_a)\cdot\left[\mathbf r(\theta_V, \Phi)-\mathbf r_0(\Phi_a) \right]=0,
\label{eq:phia}
\end{equation}
as required by the form of $\mathbf r$ in \cref{eq:posMercier}.
The angle $\theta$, present in the radius vector $\mathbf r$ in \cref{eq:posMercier}, is found using
\begin{equation}
    \theta = \arctan \left\{\frac{\left[\mathbf r(\theta_V, \Phi) - \mathbf r_0(\phi_a)\right]\cdot \mathbf b(\Phi_a)}{\left[\mathbf r(\theta_V, \Phi) - \mathbf r_0(\phi_a)\right]\cdot \mathbf n(\Phi_a)}\right\}.
\label{eq:thetamercier}
\end{equation}
The functions $\Phi_a(\theta_V, \Phi)$ and $\theta(\theta_V, \Phi)$ allow us to write the surfaces of constant flux in \cref{eq:psi02,eq:psi03} in terms of VMEC's coordinates $\theta_V$ and $\Phi$.
The data points for the fit are obtained by forming a two-dimensional grid of $\rho(\theta_V, \Phi)$ with $0\le \theta_V < 2\pi$ and $0 \le \Phi < 2\pi/N_{fp}$ with $N_{fp}$ the number of field periods of the toroidal surface ($N_{fp}=5$ for W7-X).
For this study, a total of $20\times30$ points in $(\theta_V, \Phi)$ is used.
Finally, the function $\rho(\theta_V, \Phi)$ is obtained by summing the squares of the normal and binormal components of the vector $\mathbf r - \mathbf r_0$ in \cref{eq:posMercier}, i.e.,
\begin{equation}
    \rho = \left|(\mathbf r(\theta_V, \Phi) - \mathbf r_0(\Phi_a)\right|.
\label{eq:rhovmec}
\end{equation}

The best-fit results for the Fourier coefficients of $B_0, \mu$ and $\delta$ are shown in \cref{tab:w7xfit}, where we write $B_0=\sum_n B_{0n} \cos n N_{fp} \Phi$, $\mu=\sum_n \mu_{n} \cos n N_{fp} \Phi$ and $\delta=- N_{fp} \Phi /2+\sum_n \delta_{n} \sin n N_{fp} \Phi$ with $N_{fp}=5$ for the case of W7-X.
With the functions $\delta$ and $\mu$ from \cref{tab:w7xfit}, we can estimate the rotational transform on-axis $\iota_0$ using \cref{eq:iota0}.
This yields $\iota_0=0.851$, while the rotational transform on-axis for the W7-X configuration considered here is $\iota_0=0.855$.
%

For the next order in $\psi$, {where triangularity is added as a degree of freedom}, a similar method is used to find the parameters $ \psi_{31}^{0c}, \psi_{31}^{0s}, \psi_{33}^{0c}$ and $\psi_{33}^{0s}$ that provide the best-fit results of \cref{eq:psi03} to \cref{eq:rhovmec}.
In order to make the stellarator symmetry apparent, we write $\psi_3$ as
\begin{equation}
    \psi_3 = \sum_{m,n}\psi_{3}^{mn}\cos (m \theta - n N_{fp} \Phi_a),
\end{equation}
with $N_{fp}=5$ for the case of W7-X.
The resulting Fourier coefficients $\psi_{3}^{mn}$ are shown in \cref{tab:w7xfit}, where a total of 6 Fourier modes are used.
Due to their negligible variation compared with the lowest order fit, the coefficients of $B_0, \delta$ and $\mu$ coefficients resulting from the next order fit are not shown in \cref{tab:w7xfit}.
\begin{table}
\centering
\begin{tabular}{l|llllll}
n  & 0      & 1      & 2     & 3     & 4  &  5    \\ \hline
$B_0$ & 2.78 & 0.12 &  0.01 & - &- & - \\
$\delta$ & - & 0.56  & -0.12 & 0.03 & - & - \\
$\mu$ & 0.69 & 0.20 & -0.03 & - & - & - \\
$\psi_3^{1n}$ & -0.23 & -0.45 & 0.59 & 0.12 & -0.10 & 0.15 \\
$\psi_3^{3n}$ & 0.81  & 0.42  & 0.05 & 0.61 & -0.35 & 0.23
\end{tabular}
\caption{\label{tab:w7xfit} Fitting results of the W7-X surface $\psi = 0.01$ T m$^{-2}$ to the expressions in \cref{eq:psi02,eq:psi03}. Only the parameters with absolute value greater than $0.01$ are shown.}
\end{table}

\begin{figure}
    \centering
    \includegraphics[width=0.49\textwidth]{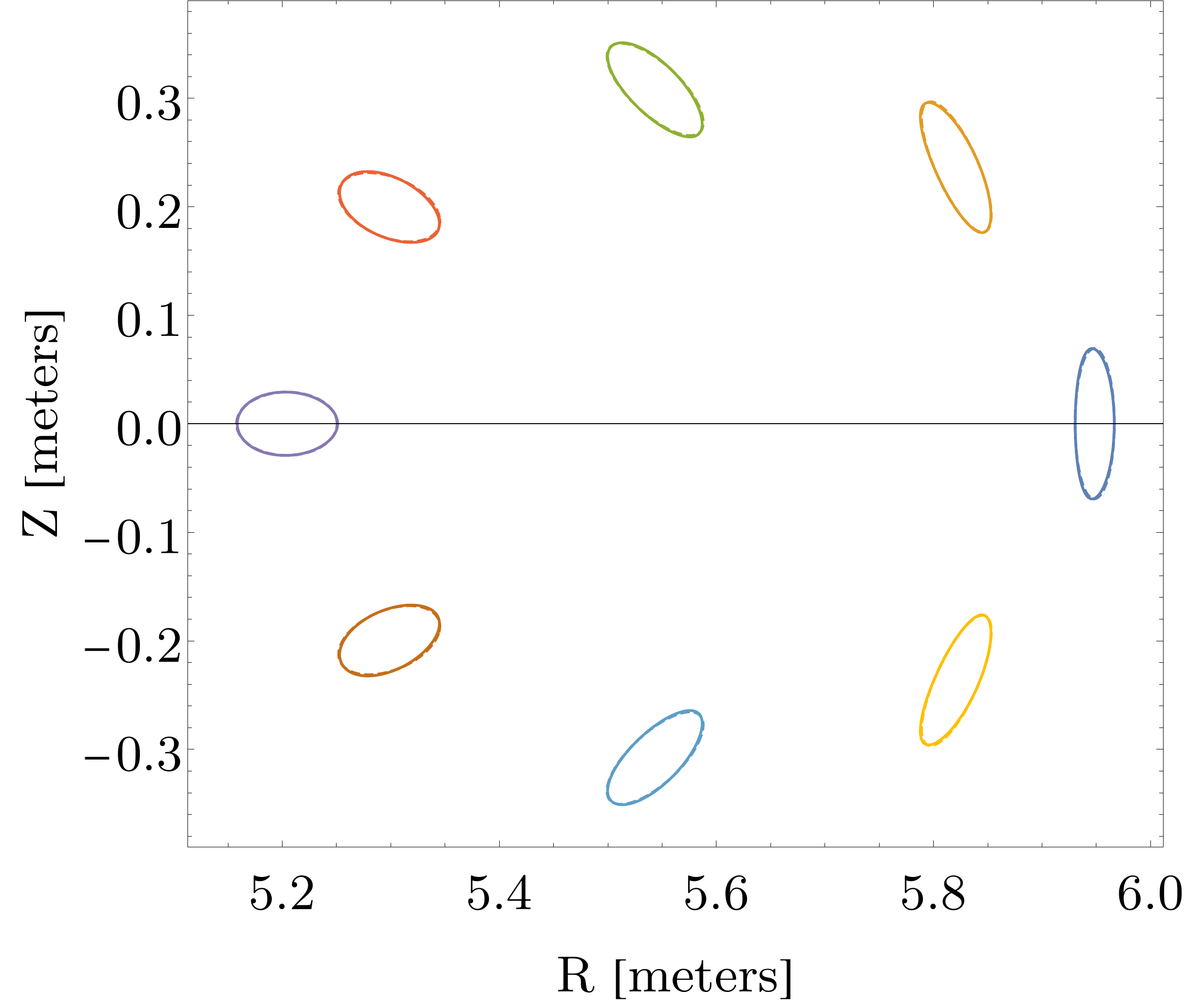}
    \includegraphics[width=0.49\textwidth]{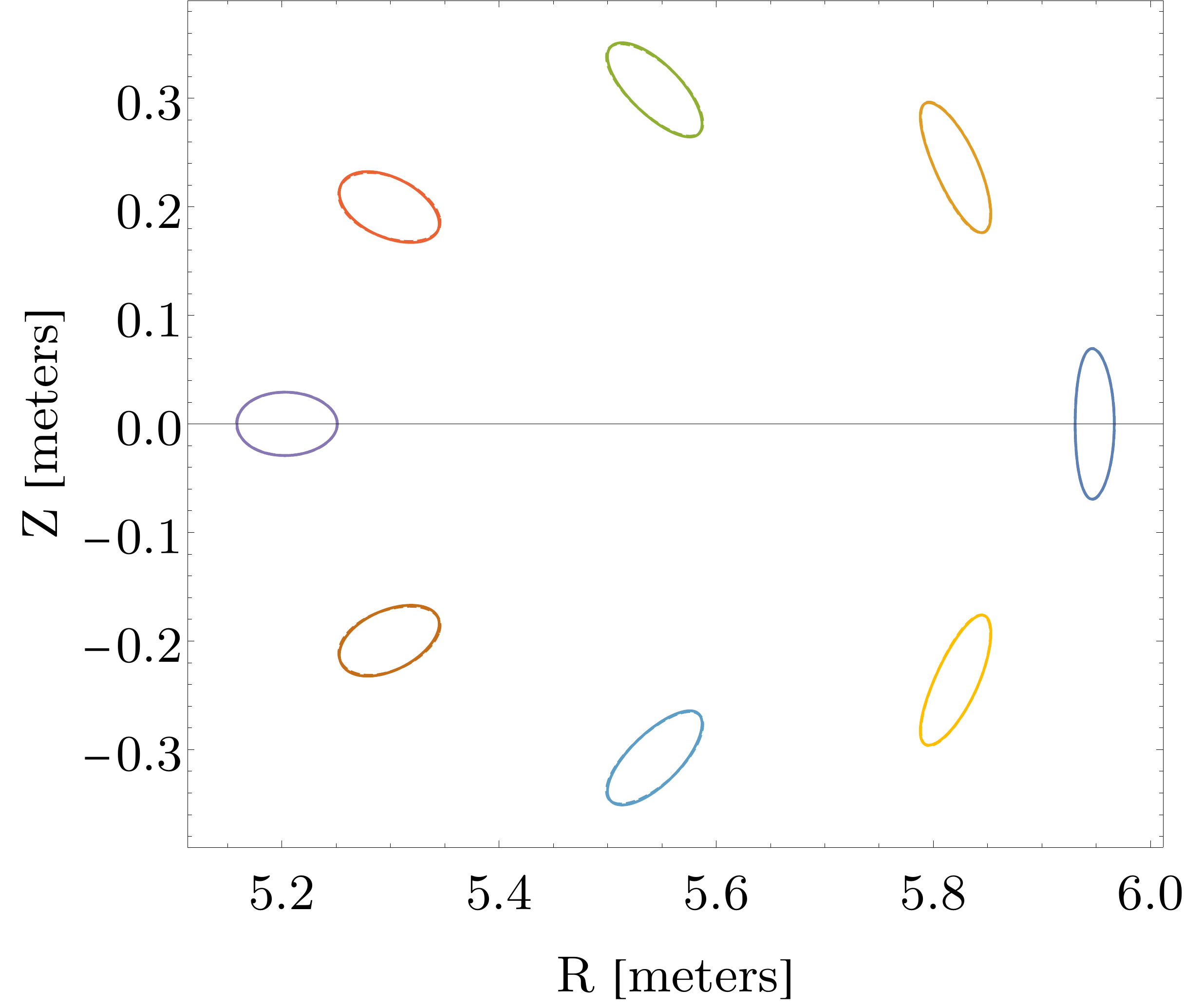}
    \includegraphics[width=0.49\textwidth]{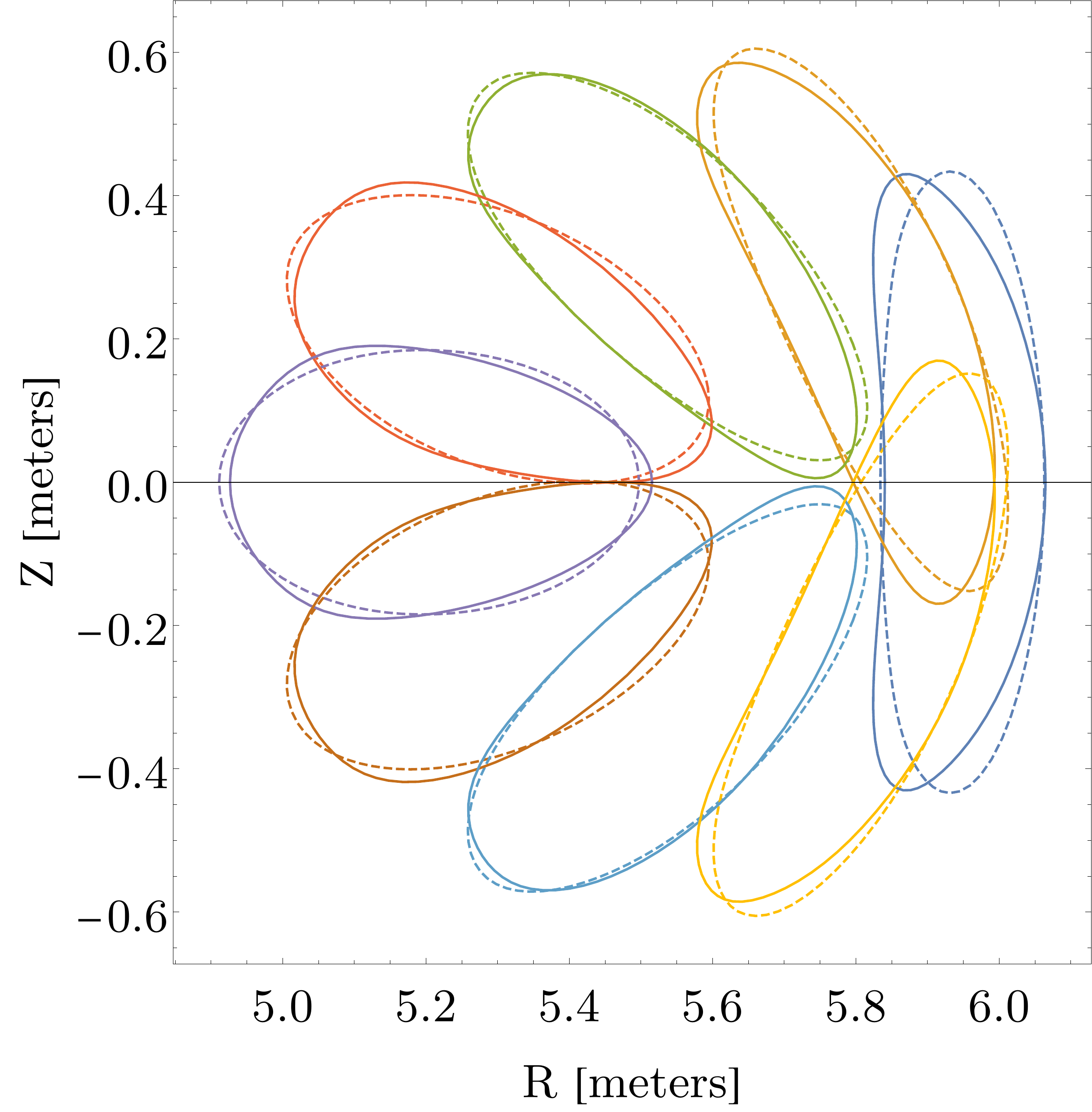}
    \includegraphics[width=0.49\textwidth]{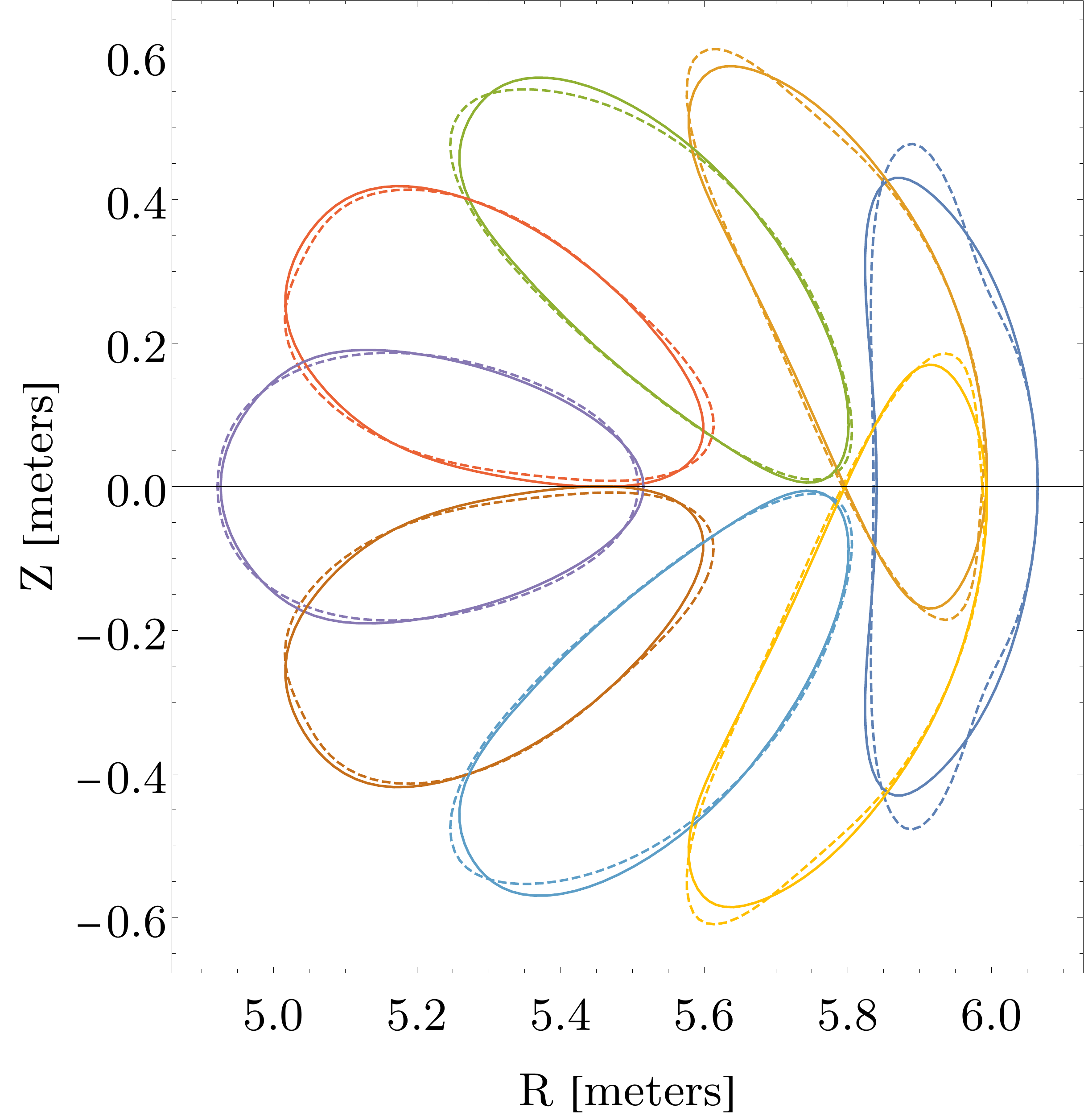}
    \caption{Eight equally spaced cross-sections with $0 \le \Phi < 2\pi/5$. Left: VMEC (full lines) and best-fit results (dashed lines) for the lowest order expression for $\psi$ in \cref{eq:psi02}. Right: VMEC (full lines) and best-fit results (dashed lines) for the higher order expression for $\psi$ in \cref{eq:psi03}. Top: fit to  surface with $\psi = 0.01$ T m$^{-2}$. Bottom: fit to surface with $\psi = 0.44$ T m$^{-2}$.}
    \label{fig:polplanes}
\end{figure}

\begin{figure}
    \centering
    \includegraphics[width=0.45\textwidth]{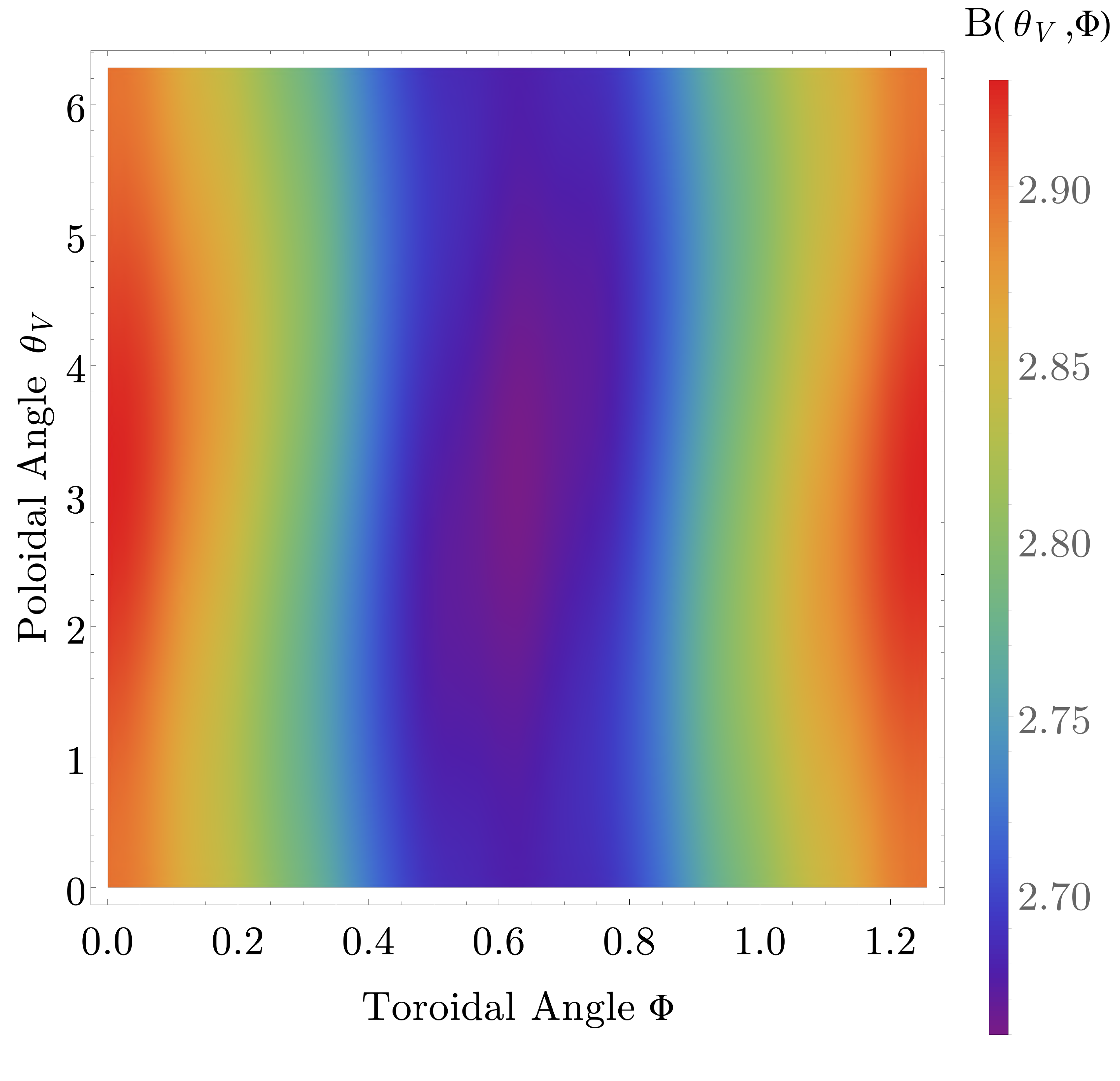}
    \includegraphics[width=0.45\textwidth]{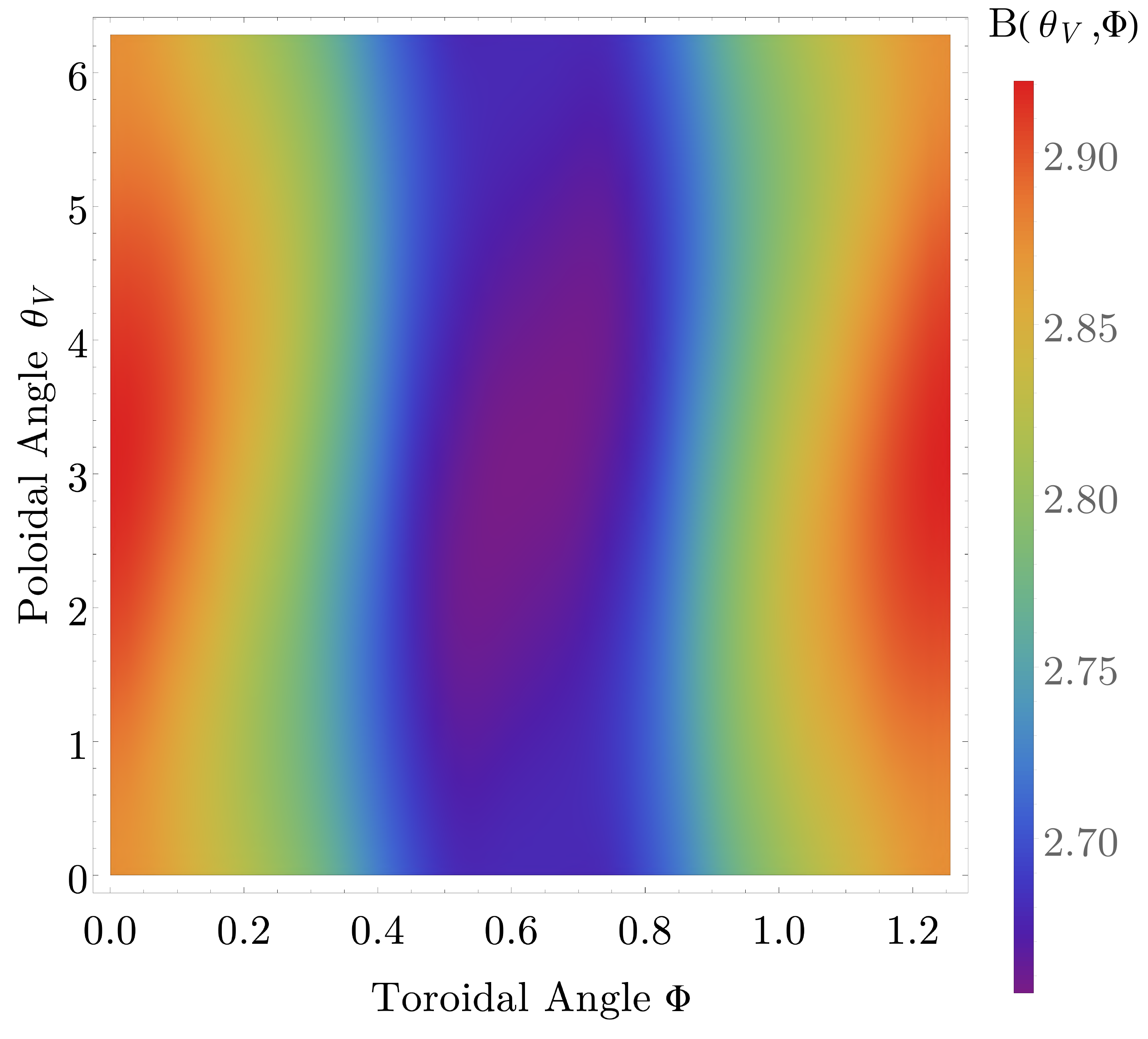}
    \includegraphics[width=0.45\textwidth]{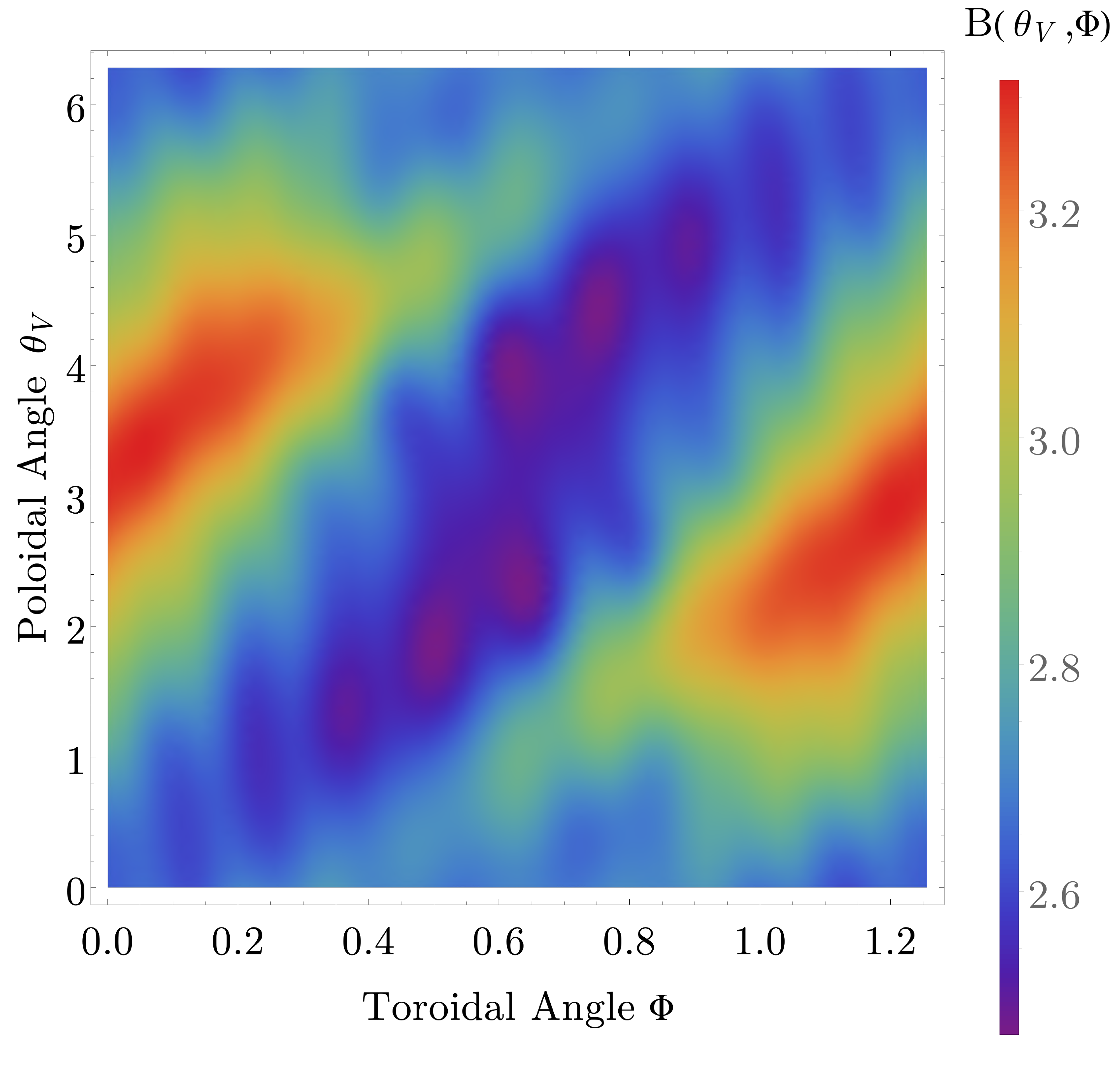}
    \includegraphics[width=0.45\textwidth]{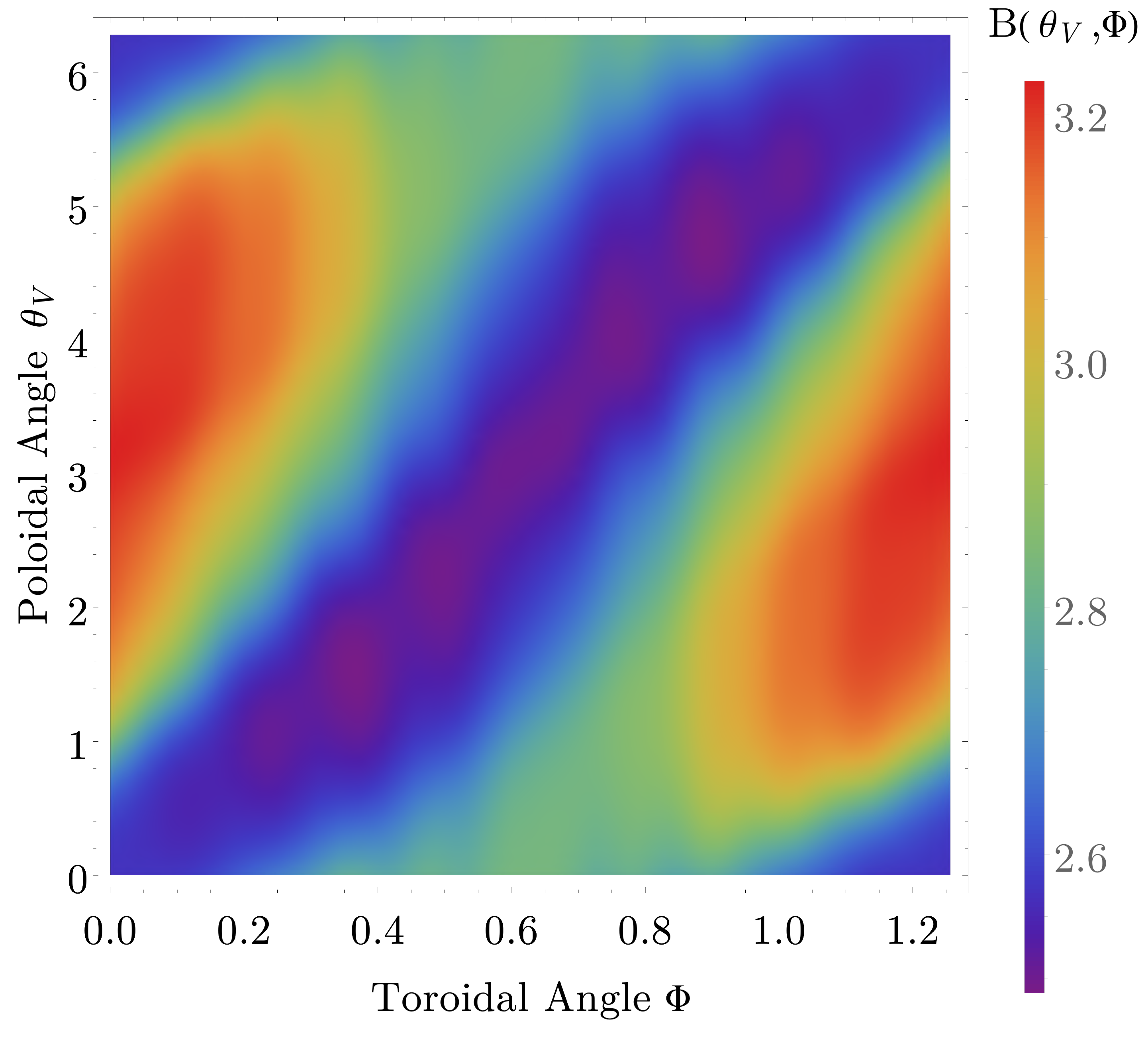}
    \caption{Magnetic field strength in the flux surface from VMEC (left) and from the lowest order fit (right) using the first order expression for $B=|\nabla \phi|$ in \cref{eq:B0firstorder}. Top: inner surface with $\psi=0.01$ T m$^2$. Bottom: Plasma boundary with $\psi=2.19$ T m$^2$.}
    \label{fig:bfieldinneroutersurf}
\end{figure}

In \cref{fig:polplanes}, we show the cross-sections of the flux surface of VMEC and the resulting lowest order (left) and higher order (right) fit results.
The eight cross-sections in \cref{fig:polplanes} are computed at equally spaced values of $\Phi$ in the interval $0 \le \Phi < 2\pi/5$.
The full lines \cref{fig:polplanes} represent VMEC's cross-sections, while the dashed lines represent the best-fit results.
Next, we compare the magnetic field on the inner W7-X surface in using the first order expression for $B=|\nabla \phi|$
\begin{equation}
    B \simeq B_0(1+\kappa \rho \cos \theta).
\label{eq:B0firstorder}
\end{equation}
In \cref{fig:bfieldinneroutersurf} (top) we show the magnetic field strength in the inner surface from VMEC (left) and from the lowest order fit (right using the first order expression for $B$ in \cref{eq:B0firstorder}, while in \cref{fig:bfieldinneroutersurf} (bottom) the same is shown for the plasma boundary surface.

\begin{figure}
\begin{minipage}{0.47\textwidth}
\begin{subfigure}{\textwidth}
\includegraphics[width=\linewidth]{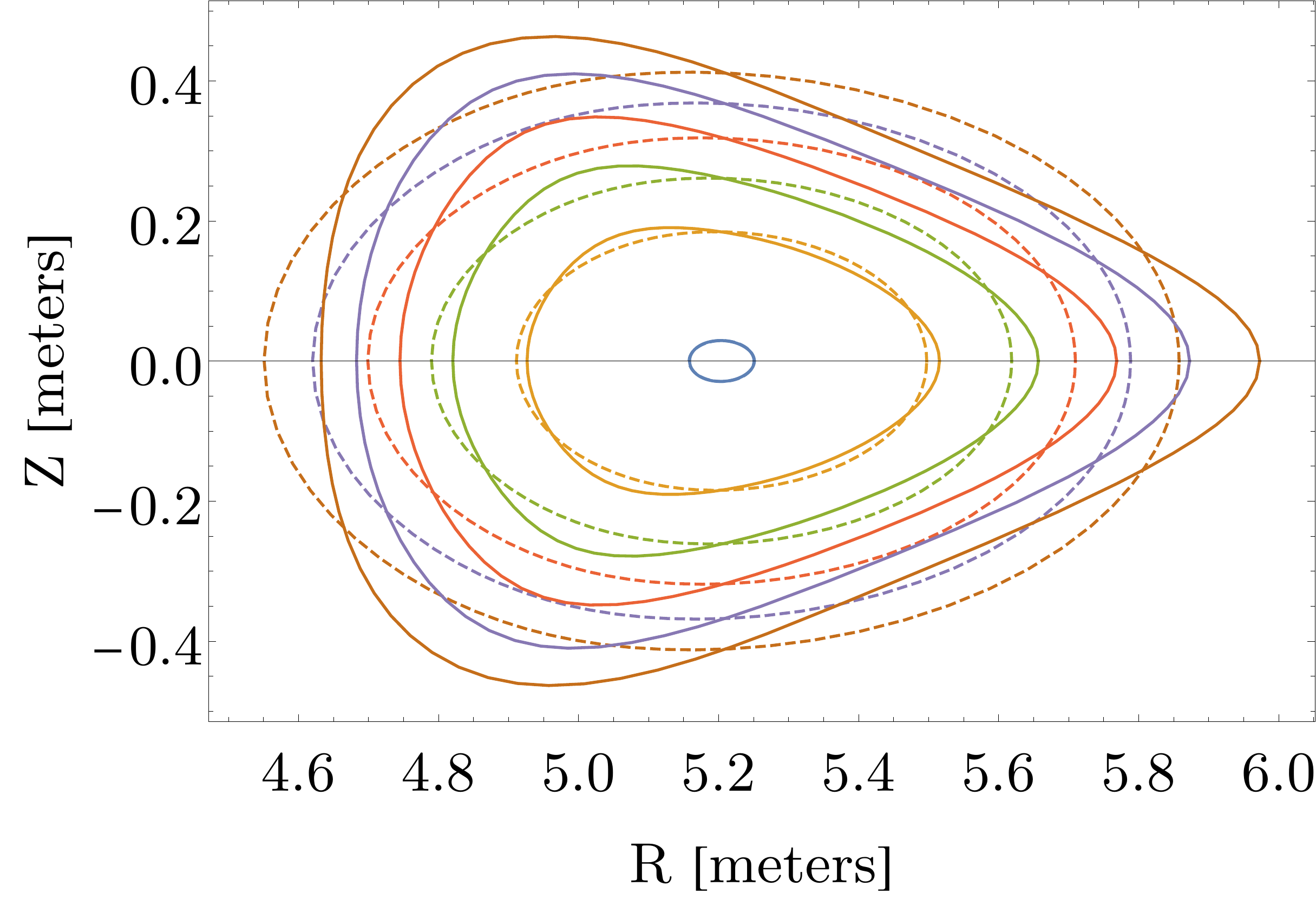}
\end{subfigure}
\hfill
\begin{subfigure}{\textwidth}
\includegraphics[width=\linewidth]{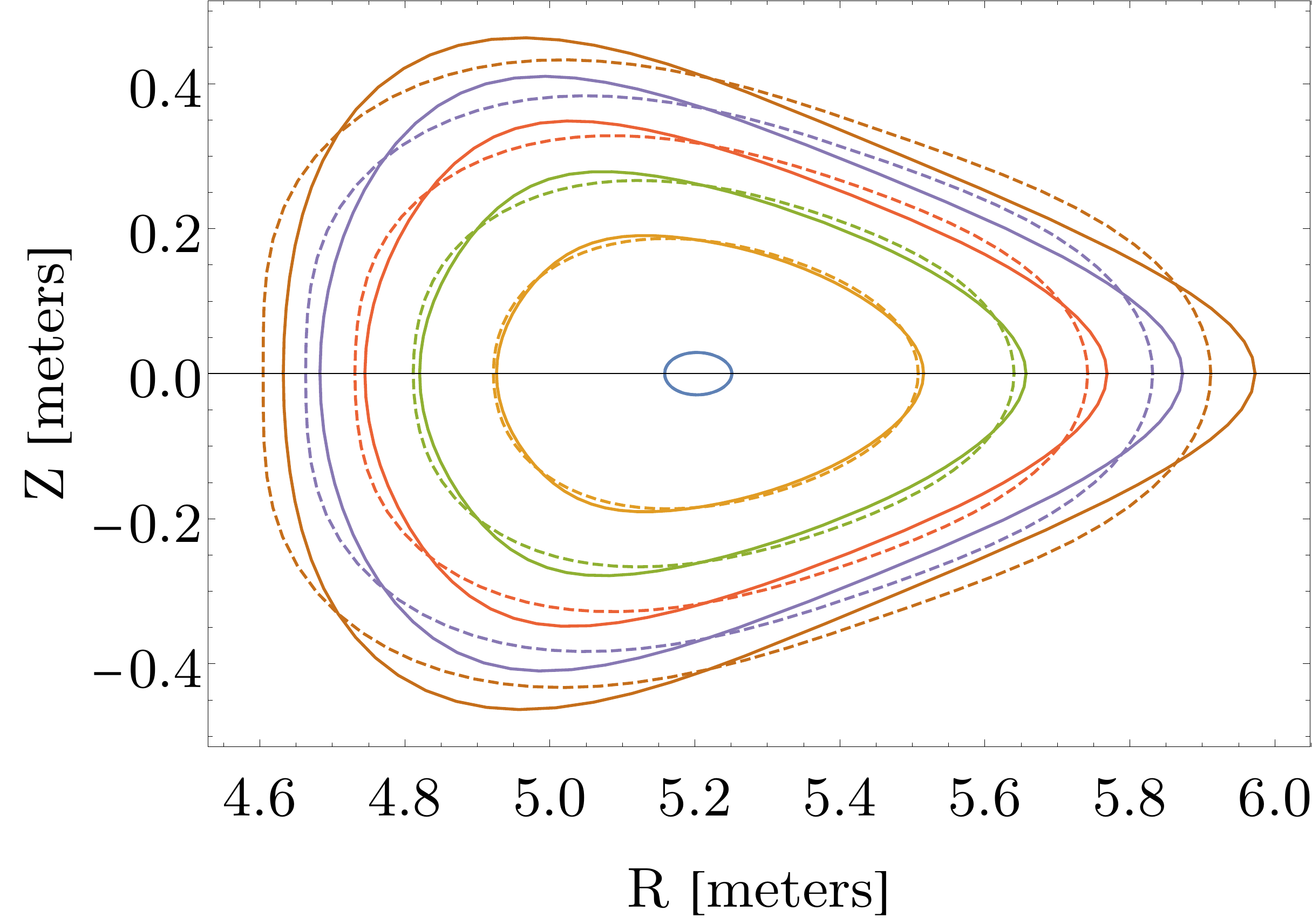}
\end{subfigure}
\end{minipage}
\begin{subfigure}{0.52\textwidth}
\includegraphics[width=\linewidth]{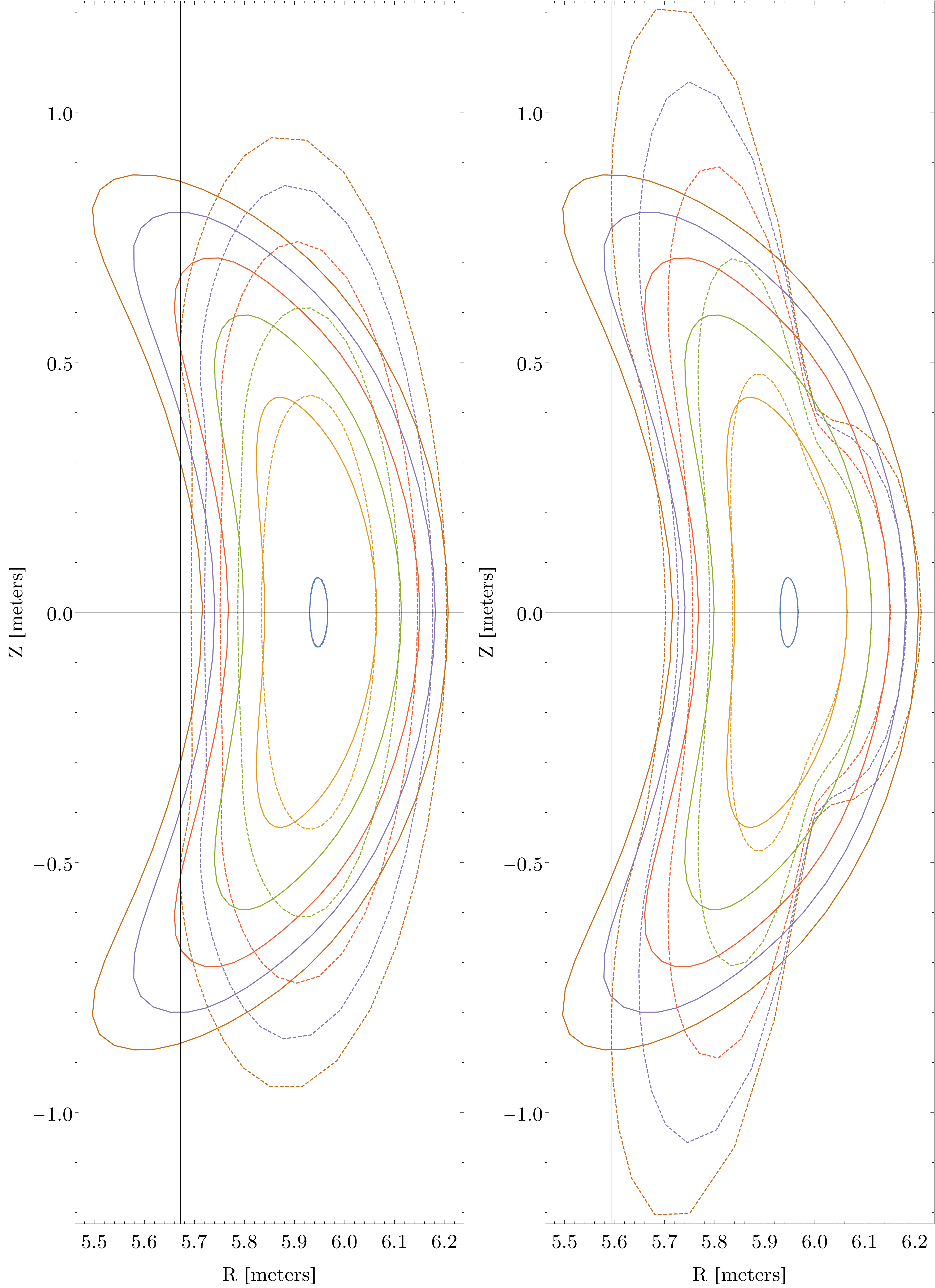}
\end{subfigure}
\caption{\label{fig:sixsurf1}
Left: cross-section of VMEC (full lines) and the best-fit results (dashed lines) at lowest order (top) and higher order (bottom) at $\Phi= 2\pi/10$; Right: cross-section of VMEC (full lines) and the best-fit results (dashed lines) at lowest order (left) and higher order (right) $\Phi= 0$.}
\end{figure}

\begin{figure}
    \centering
    \includegraphics[width=0.45\textwidth]{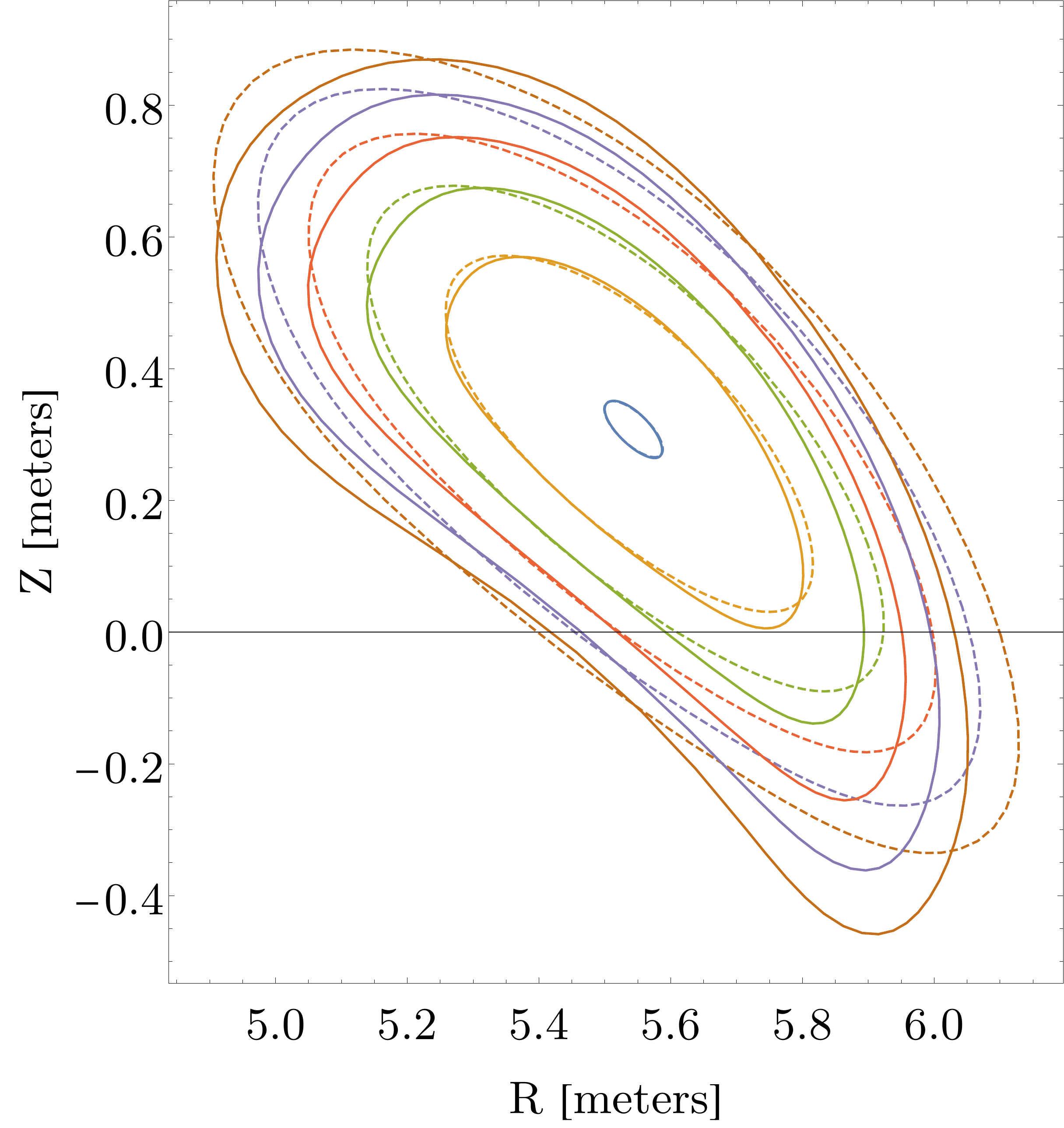}
    \includegraphics[width=0.45\textwidth]{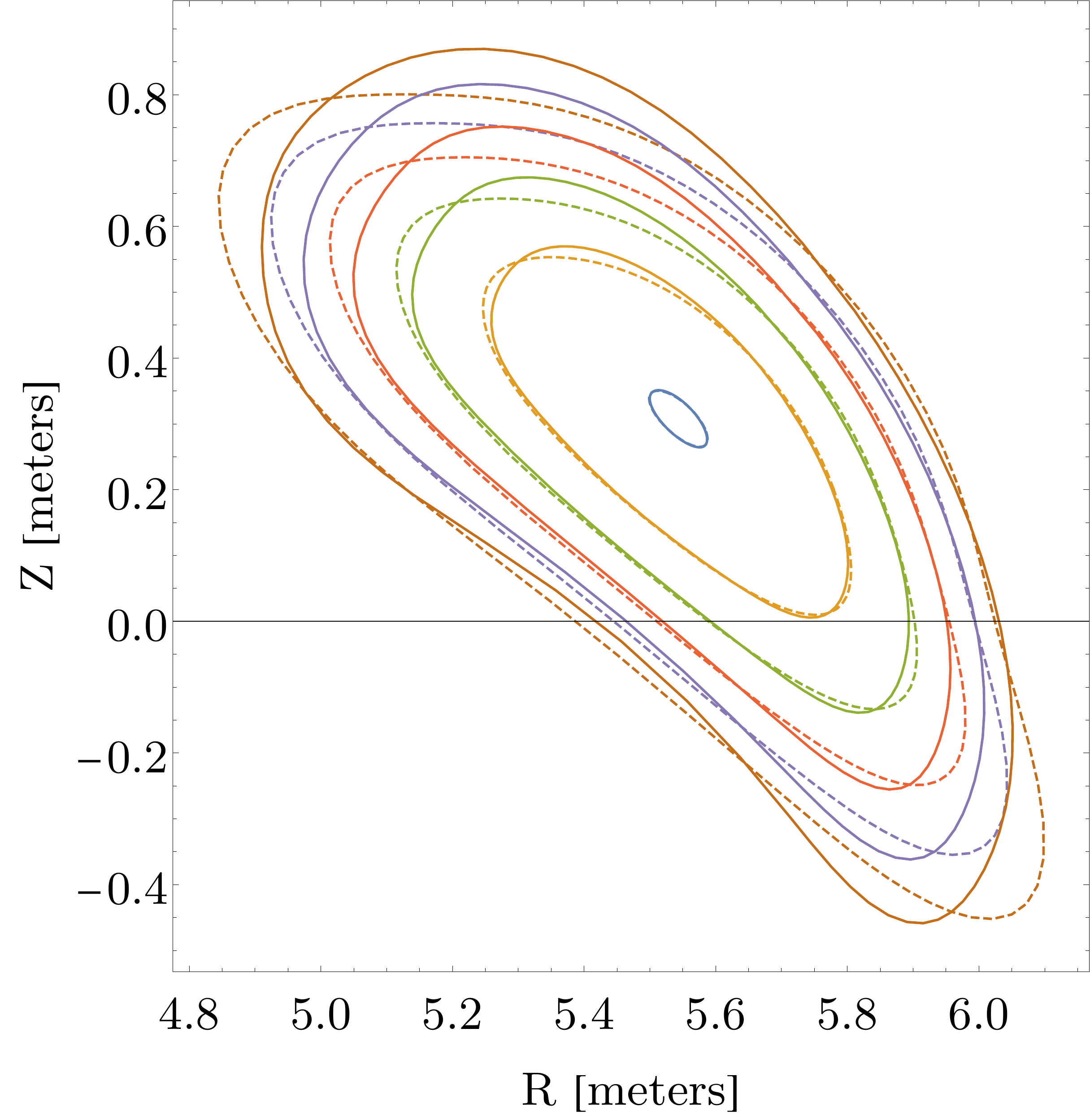}
    \caption{Cross-section of VMEC (full lines) and the best-fit results (dashed lines) at lowest order (left) and higher order (right) at $\Phi= \pi/10$.}
    \label{fig:sixsurf2}
\end{figure}

\begin{figure}
    \centering
    \includegraphics[trim={4cm 3.5cm 3.5cm 2cm}, clip,width=0.49\textwidth]{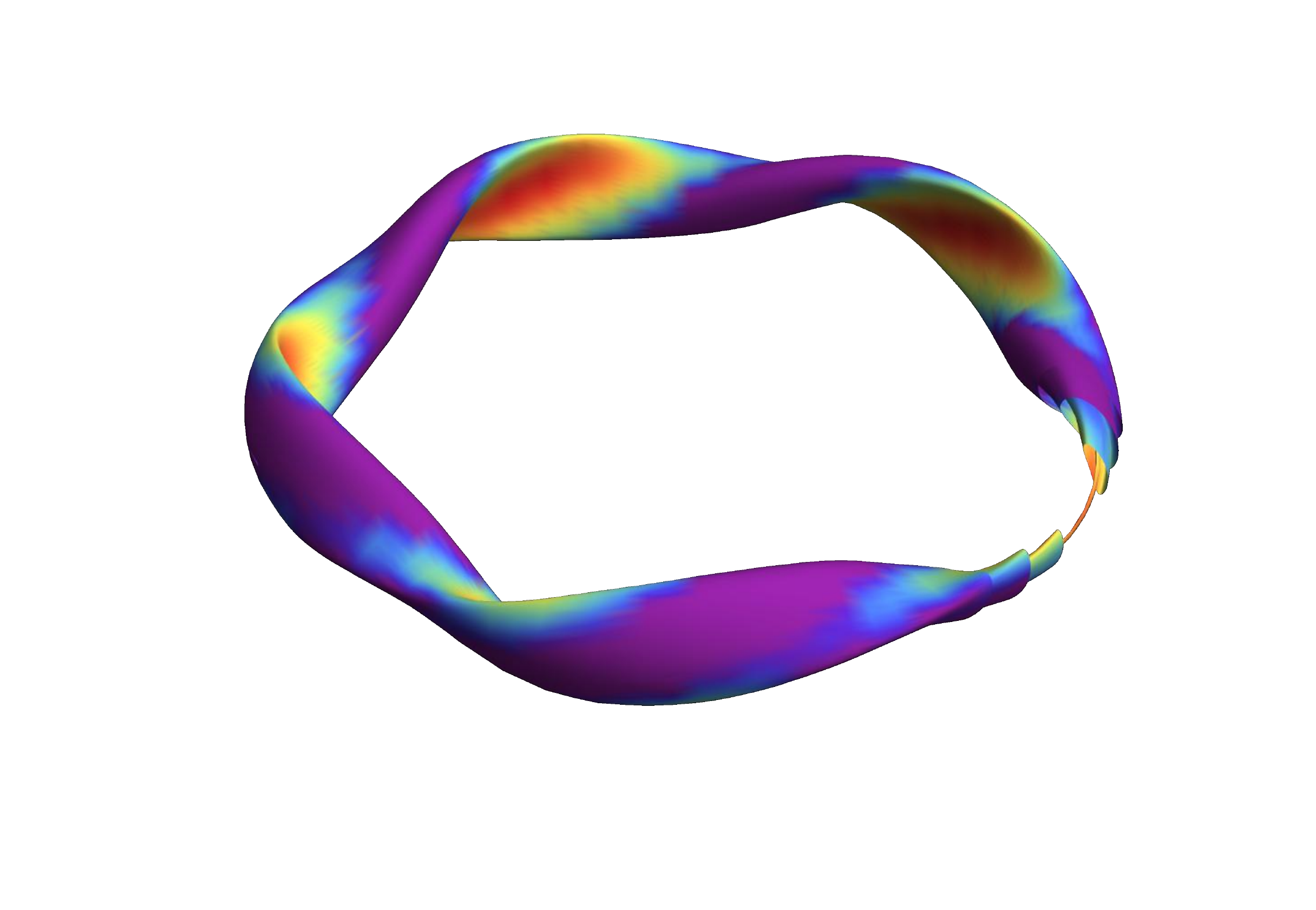}
    \includegraphics[trim={4cm 3.5cm 3.5cm 2cm}, clip,width=0.49\textwidth]{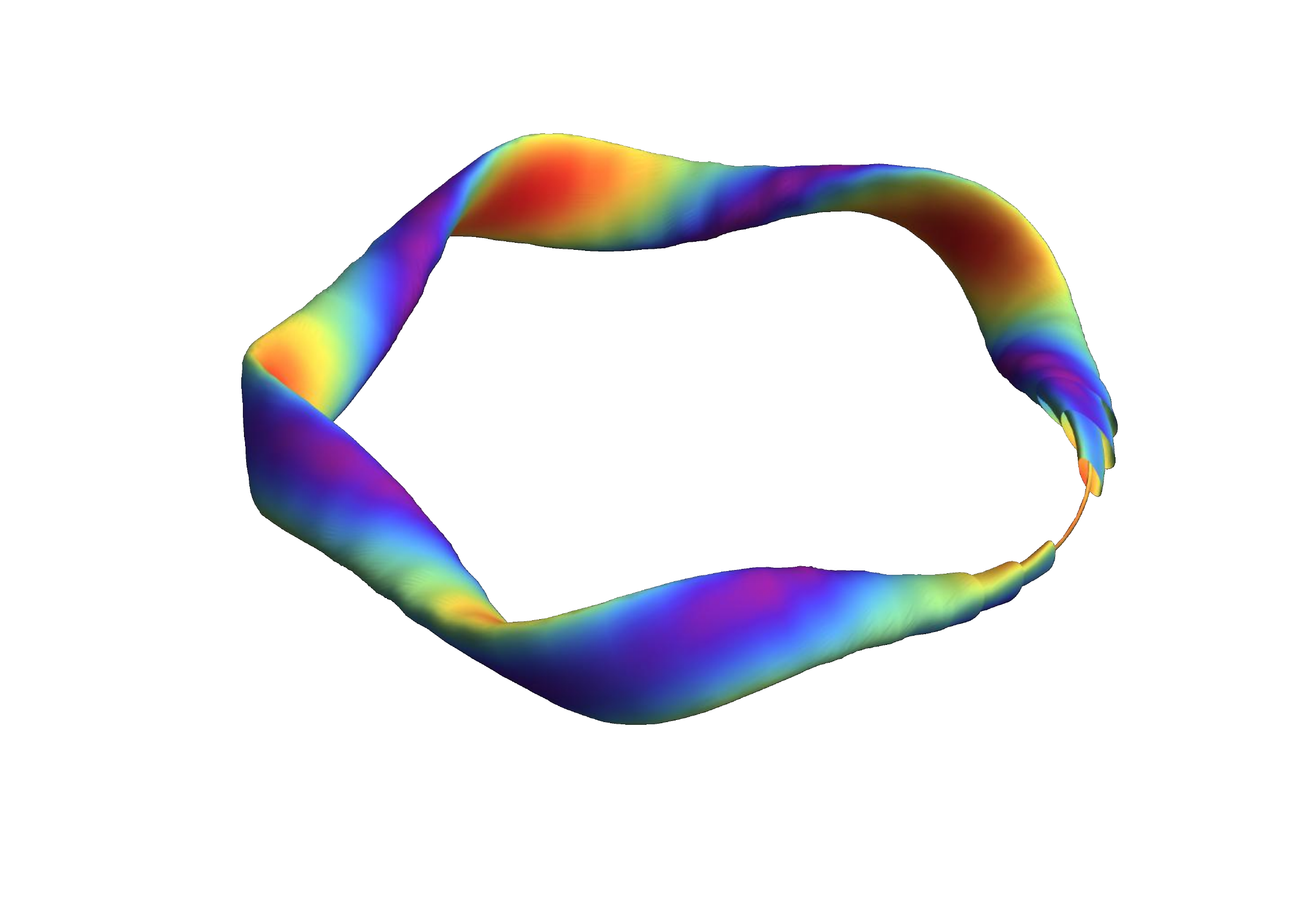}
    \caption{Left: W7-X plasma boundary. Right: Resulting surface from the nonlinear squares fit of \cref{eq:psi02} to \cref{eq:rhovmec}. The colors show the the magnetic field strength on the surface.}
    \label{fig:3DsurfacesPB}
\end{figure}

Finally, we look at the cross-sections of six equally-spaced surfaces of constant $\psi$ from the inner surface to the plasma boundary using the best-fit results of the nonlinear regression to the inner surface obtained in \cref{tab:w7xfit}.
In \cref{fig:sixsurf1}, we show the cross-section of VMEC (full lines) and the best-fit results (dashed lines) at lowest order (left) and higher order (right) at $\Phi =0$ and $\Phi = \pi/5$, while in \cref{fig:sixsurf2} we show the cross-sections at $\Phi=\pi/10$.
We note that the parameters obtained for the considered inner surface where $\epsilon \rho=10^{-4}$ yield a shaping of the surfaces up to the plasma boundary that follow qualitatively the behaviour of the flux surfaces obtained using the VMEC code except at $\Phi=0$, where the agreement is limited to the inner surfaces.
This is also seen in \cref{fig:3DsurfacesPB}, where the plasma boundary of W7-X and its magnetic field strength is compared to the resulting surface using the best-fit parameters for $\psi_2$.


In this section, we were able to obtain both second and third order approximations in $\epsilon$ for the surfaces of constant toroidal flux $\psi$ of the W7-X stellarator by performing a nonlinear regression to a single surface close to the magnetic axis, which requires very little computational effort to compute.
This procedure has shown to yield the correct rotational-transform on-axis with an error of less than 0.5\%, and to predict the qualitative behavior of the shape and strength of the magnetic field across a wide range of volumes inside the plasma boundary.
We remark that the method described here is valid at arbitrary order for a tokamak or stellarator-like toroidal equilibrium obtained using the VMEC code.

\section{Conclusion}

In this work, equilibrium magnetic fields are constructed at arbitrary order in the distance from the magnetic axis to the outer plasma boundary, both for vacuum and finite-$\beta$ configurations.
Using an orthogonal coordinate system based on the parameters of the magnetic axis, the coefficients of the asymptotic power series in $\epsilon$ (the inverse aspect ratio) for the magnetic field, magnetic flux surface function, field line label, and rotational transform are derived.
While the near-axis framework allows for the construction of magnetic fields with  chaotic structure, it also allows for the existence of nested flux surfaces.
The associated constraints for the non-existence of good flux surfaces are derived, namely the presence of resonant denominators that vanish for a rational rotational transform.
Within a finite-$\beta$ construction, a procedure to compute the resulting Shafranov shift of the magnetic axis and the associated $\beta$ limit is presented, and a comparison between the lowest order direct and inverse coordinate methods is shown.
Finally, a numerical analysis is performed by comparing the near-axis expansion to a W7-X equilibrium at second and third order in the expansion.

The framework presented here is applicable to a wide range of plasma configurations.
Indeed, as shown in \citet{Landreman2019}, the lowest order inverse coordinate approach (which is shown in this work to be equivalent to the direct approach used here), when used to construct quasisymmetric designs, can accurately describe many stellarator designs obtained using numerical optimization algorithms.
The construction of quasisymmetric stellarator shapes using the methods developed here will be the subject of future work.
As a further avenue of future study, we mention the possibility of using the near-axis expansion to numerically compute stellarator shapes in the volume inside a given surface by solving the system of equations in \cref{eq:odepsi0n} to compute new equilibria as opposed to the nonlinear regression approach applied here.
Furthermore, a thorough study of the resonances present in the system of \cref{eq:brho1eq,eq:bomega1eq,eq:bs1eq} that might lead to the appearance of non-analytic and weakly singular terms of the form $\rho^n (\log \rho)^m$ is needed.
Finally, by using a sufficiently high order in the expansion, we expect this framework to be able to generate input data for optimization codes such as ROSE \citep{Drevlak2019} and STELLOPT \citep{Spong2001}, possibly increasing the performance of such numerical tools.

\section{Acknowledgements}

We wish to thank J. Loizu, A. Cerfon, P. Helander, G. Plunk, R. MacKay and N. Kalinikos for many fruitful discussions regarding the content of the manuscript, as well as the Courant Institute, IPP Greifswald and the University of Warwick for their warm reception during the writing of the manuscript.
This work was supported by a grant from the Simons Foundation (560651, ML) and a US DOE Grant No. DEFG02-86ER53223.

\appendix

\section{Fourier expansion of $f_n(\omega,s)$}
\label{app:Fourierphi}

The aim of this section is to show that the magnetic scalar potential $f_n$ in \cref{eq:phirhoomega} can be written as a Fourier series with frequencies $p$ ranging from $0 \le p \le n$ or, equivalently, to derive \cref{eq:phinp}.
We start by deriving the Fourier series using \cref{eq:analyticity}, which stems from an expansion in $x$ and $y$, and then perform a similar derivation instead using Laplace's equation, \cref{eq:arbitraryorderphi}, which stems from an expansion in $\rho$.

Starting with \cref{eq:analyticity}, we aim at deriving the Fourier series of the product $\cos^p{\theta}\sin^{(n-p)}\theta$.
While the expressions for the Fourier series of $\sin^n{ \theta}$ and $\cos^n{\theta}$ can be found in previous literature \citep{Zwillinger2014}, a brief derivation for the Fourier series of their product is given below.
Such product can be simplified using Euler's formula and the binomial theorem, yielding
\begin{align}
    \cos^p{\theta}\sin^{(n-p)}{\theta}&=\frac{1}{2^n}\frac{1}{i^{(n-p)}}\sum_{r=0}^p \sum_{k=0}^{n-p}\binom{p}{r}\binom{n-p}{k}(-1)^k e^{i \theta[(n-2(k+r)]}\nonumber\\
    &=\frac{1}{2^n}\frac{1}{i^{(n-p)}}\sum_{r=0}^p \sum_{\nu=r}^{n-p+r}\binom{p}{r}\binom{n-p}{\nu-r}(-1)^{\nu-r} e^{i \theta (n-2\nu)}, \label{eq:doubleFourier}
\end{align}
where, in the last step, we replaced the summation index $k$ with $\nu=k+r$.

We now simplify \cref{eq:doubleFourier} by interchanging the sum limits and split the result between even and odd $n$.
For odd $n$, the right-hand side of (\ref{eq:doubleFourier}) is given by
\begin{align}
 \frac{1}{2^n}\frac{1}{i^{(n-p)}}\left[\sum_{\nu=0}^p \left(e^{i \theta (n-2\nu)}+(-1)^{n-p}e^{-i \theta (n-2\nu)}\right)\sum_{r=0}^{\nu}\binom{p}{r}\binom{n-p}{\nu-r}(-1)^{\nu-r} \right. \nonumber\\
 \left.+\sum_{\nu=p+1}^{(n-1)/2}  \left(e^{i \theta (n-2\nu)}+(-1)^{n-p}e^{-i \theta (n-2\nu)}\right)\sum_{r=0}^{p}\binom{p}{r}\binom{n-p}{\nu-r}(-1)^{\nu-r}\right],
\end{align}
with the terms involving the exponential function reducing to $\sin{(n-2\nu)\theta}$ or $\cos{(n-2\nu)\theta}$ if $n-p$ is odd or even, respectively.
The sums over $r$ can be stated in terms of the Gaussian hypergeometric function ${_2}F_1$, yielding
\begin{align}
    \sum_{r=0}^{\nu}\binom{p}{r}\binom{n-p}{\nu-r}(-1)^{\nu-r}&=c(\nu,n,p)\nonumber\\
    &=\sum_{r=0}^{p}\binom{p}{r}\binom{n-p}{\nu-r}(-1)^{\nu-r} \nonumber\\
    &=(-1)^\nu \binom{n-p}{\nu}{_2}F_1(-p,-\nu;1+n-p-\nu;-1)\\
    &=(-1)^\nu \frac{\Gamma(1+n-p)}{\Gamma(1+\nu)}\frac{{_2}F_1(-p,-\nu;1+n-p-\nu;-1)}{\Gamma(1+n-p-\nu)}.
\end{align}
The last identity shows that $c(\nu,n,p)$ is well defined even when $(1+n-p-\nu)$ is negative \citep{Olver2010}. Therefore, for odd $n$, the product in \cref{eq:doubleFourier} can be written as
\begin{align}
    \cos^p{\theta}\sin^{(n-p)}{\theta}&=2^{n-1}(-1)^{\frac{n-p-1}{2}}\sum_{\nu=0}^{(n-1)/2} c(\nu,n,p) \sin{(n-2\nu)\theta}\quad \text{(for even } p),\\
    &=2^{n-1}(-1)^{\frac{n-p}{2}}\sum_{\nu=0}^{(n-1)/2} c(\nu,n,p)\cos{(n-2\nu)\theta}\quad \text{(for odd } p).
\label{eq:cospsinnp}
\end{align}
We note that for odd $n$, the range of the Fourier modes are from $\theta$ [when $\nu=(n-1)/2$] to $n$ (when $\nu=0$), and only odd harmonics are present. 

The results for even $n$ differ from the above due to the additional $\nu=n/2$ term. Proceeding as before, we write the right-hand side of $\cref{eq:doubleFourier}$ as
\begin{align}
    &{2^{-n}}{(-1)^{\frac{n-p-1}{2}}} [c\left(n/2,n,p\right)+\sum_{\nu=0}^{(n/2)-1}2c(\nu,n,p) \sin{(n-2\nu)\theta}]
\end{align}
for odd $p$, while for even $p$ it reads
\begin{align}
    {2^{-n}}(-1)^{\frac{n-p}{2}} [c\left(n/2,n,p\right)+\sum_{\nu=0}^{(n/2)-1}2c(\nu,n,p) \cos{(n-2\nu)\theta}].
\end{align}
In this case, the Fourier modes lie between $0$ and $n$, and only even modes appear. %
%
As shown above, there are no Fourier modes with frequency higher than $n$ in \cref{eq:doubleFourier}, showing that $f_n$ is indeed analytic.

We now show the analyticity of $f_n$ using Laplace's equation, \cref{eq:arbitraryorderphi} as a starting point, i.e., derive \cref{eq:phinp} from \cref{eq:arbitraryorderphi}.
We start by Fourier decomposing use the $\cos^n{\theta}$ term in \cref{eq:arbitraryorderphi}, yielding \citep{Zwillinger2014} 
\begin{equation}
\cos^n{\theta}= \sum_{\nu=0}^{\lfloor n/2\rfloor}C_{e}(\nu,n)\cos{(n-2\nu)\theta}\label{eq:cospowern}
\end{equation}
with $C_{e}(\nu,n)= {2^{1-n}}\binom{n}{\nu}$ for odd $n$ and $C_{e}(\nu,n)={2^{-n}}(2\binom{n}{\nu}+\binom{n}{n/2}\delta_{\nu,n/2})$ for even $n$.
We note that by replacing the index $p$ with $n$ in the Fourier expansion of $\cos^p{\theta}\sin^{(n-p)}{\theta}$ of \cref{eq:cospsinnp}, the expression in \cref{eq:cospowern} is obtained.
Using \cref{eq:cospowern}, the right-hand side of (\ref{eq:arbitraryorderphi}) can be rewritten as
\begin{align}
    \sum_{m=0}^{n-3}\sum_{\nu=0}^{\lfloor m/2\rfloor}&C_{e}(\nu,m)\cos{(m-2\nu)\theta}\left[\kappa\left(\dot f_{n-m-1} \sin \theta-(n-m-1)f_{n-m-1}\cos \theta\right)\right.\nonumber\\
    &\left.+(m+1)f_{n-m-2}''+\frac{(m+1)(m+2)}{2}(\kappa' \cos \theta+\kappa \tau \sin{\theta}) f_{n-m-3}'\right] \kappa^m.
\label{eq:arbit}
\end{align}
Splitting \cref{eq:arbit} between its $(m+1)$ and $(m-2\nu\pm 1)$ harmonics, we obtain
\begin{align}
\sum_{m=0}^{n-3} &\sum_{\nu=0}^{\lfloor m/2\rfloor} C_{e}(\nu,m)  \kappa^m \left[\left(\cos{(m-2\nu+1)\theta}+\cos{(m-2\nu-1)\theta}\right)T_1\right.\nonumber\\
    &\left.+\left(\sin{(m-2\nu+1)\theta}+\sin{(m-2\nu-1)\theta}\right)T_2+2(m+1)f_{n-m-2}'' \cos{\theta} \right], \label{eq:arbitRHS}
\end{align}
where
\begin{align}
     T_1 &= \frac{(m+1)(m+2)}{2}\kappa' f_{n-m-3}'-(n-m-1)\kappa\: f_{n-m-1},
\label{eq:T1exp}
\end{align}
and
\begin{align}
     T_2 &=\kappa \left[\frac{(m+1)(m+2)}{2} \tau f_{n-m-3}'+\dot f_{n-m-1}\right].
\label{eq:T2exp}
\end{align}
From the  $(m-2\nu\pm 1)$ harmonic terms \cref{eq:arbitRHS}, it is clear that the maximum frequency of $\omega$ in \cref{eq:arbitraryorderphi} is $(n-2)$.
This shows that, in vacuum, the forced harmonic oscillator equation determining the magnetic field is free of resonances of frequency $n$.

We shall now inductively prove that $\phi_k$ is of the form given by (\ref{eq:phinp}) for arbitrary order in $k$.
The case for $k \le 3$ is already derived in \cref{eq:phi2eq,eq:phi3eq}.
To order $k>3$, we assume the form (\ref{eq:phinp}), such that
\begin{align}
   \phi_k &=\sum_{p=0}^k \phi_{pk}^c(s) \cos{p u}+\phi_{pk}^s(s)\sin{p u}, \nonumber\\
   \dot{\phi}_k &=\sum_{p=0}^k p\left[ -\phi_{pk}^c(s) \sin{p u}+\phi_{pk}^s(s)\cos{p u}\right],\label{eq:dotphinp}\\
    \phi'_k &=\sum_{p=0}^k \left\{ \left[{\phi_{pk}^c}'(s)-p \tau\: \phi_{pk}^s(s) \right]\cos{p u}+\left[{\phi_{pk}^s}'(s)+p \tau\: \phi_{pk}^c(s)\right]\sin{p u}\right\}. \nonumber
\end{align}
The maximum frequency of $\omega$ in $T_1$ and $T_2$ stems from the $\phi_{n-m-1}$ terms and is $(n-m-1)$.
Similarly, the term $\phi_{n-m-2}''\cos{\theta}$ yields a maximum frequency $(n-m-1)$.
When plugged in (\ref{eq:arbitRHS}), we obtain frequencies in the range $m-2\nu\pm 1-n+m+1$ to $m-2\nu\pm 1 + n-m-1$ with an upper limit of $n-2$. Therefore, the form (\ref{eq:phinp}) holds for arbitrary order in $n$.
The terms with frequency $p$ in the range $0\le p \le n-2$ in \cref{eq:phinp} for $\phi_n$ are shown to be determined by its lower order counterparts, while two free functions of $s$ are obtained at each order $n$, i.e., the $\cos$ and $\sin$ coefficients of frequency $n$.

\section{Solution for $\psi_n$ using the method of characteristics}
\label{app:characteristics}

Noting that \cref{eq:psi0n} is of the form of an inhomogeneous advection equation for $\psi^0_{n}$, its solution can either be solved iteratively using the analyticity condition, \cref{eq:analyticity}, or by using the method of characteristics.
Along the characteristic curves parameterized by $s$ with $d\omega/ds=\dot \phi_2/B_0$, we can write \cref{eq:psi0n} as
\begin{equation}
    \frac{d \psi^0_{n}}{ds}+\frac{2 n \phi_2}{B_0}\psi^0_{n}=\frac{F^0_{n}}{B_0}.
\label{eq:psi0nadv}
\end{equation}
Using the expression for $\phi_2$, the characteristic curve $\omega(s)$ is given by
\begin{equation}
    \tan u = \frac{\eta'}{2 u' \mu}+\frac{\zeta_1(s)}{u' \mu}\tanh\left[\zeta_2(s)+\zeta_1(s) B_0(s)\int_0^s\frac{ds'}{B_0(s')}\right],
\end{equation}
where $\zeta_1(s)=\sqrt{(\eta'/2)^2+\mu^2 (u')^2}$ and $\zeta_2(s)$ is an arbitrary function of $s$.
Defining the integrating factor $M_n(s)= \exp(2 n\int_{0}^{s}\phi_2(s')/B_0(s')d s')$, the solution of \cref{eq:psi0nadv} reads
\begin{equation}
    \psi^0_{n}(s)=\frac{1}{M_n(s)}\left[C+\int_{0}^{s}\frac{M_n(s')F_n(s')}{B_0(s')}d s'\right].
\end{equation}
The integration constant $C$ is found imposing the periodicity requirement $\psi^0_{n}(s+L)=\psi^0_{n}(s)$, with $L$ the total length of the magnetic axis, yielding
\begin{equation}
    C=\frac{\oint \frac{M_n F_n}{B_0}d s}{1-e^{2n\oint\frac{\phi_2}{B_0}ds}}.
\end{equation}

\section{Explicit Expressions for the Higher Order Vacuum Flux Surface Function}
\label{app:higherorder}

For the $n=3$ case in \cref{eq:psi0n}, the source term reads $F^0_3=-6 \phi_3 \psi^0_{2}-\dot \phi_3 \dot \psi^0_{2}-2 B_0 \psi^{0'}_{2} \kappa \cos \theta$.
Similarly to \cref{eq:systempsi02}, the equation for $\psi^0_{3}$, with $\psi^0_3$ of the form

\begin{equation}
    \psi^0_3=\psi^{0c}_{31} \cos u + \psi^{0s}_{31} \sin u+\psi^{0c}_{33} \cos 3u+\psi^{0s}_{33} \sin 3 u,
    \label{eq:psi03}
\end{equation}
can be written as
\begin{equation}
    \Psi^{0'}_{3}=A^0_{3} \Psi^0_{3}+B^0_{3},
\label{eq:systempsi03}
\end{equation}
with $\Psi^0_{3}=B_0^{-3/2}[\psi^{0c}_{31}  \psi^{0s}_{31} \psi^{0c}_{33} \psi^{0s}_{33}]^T$, $A^0_{3}$ the matrix
\begin{equation}
    A^0_{3}= 
\left(
\begin{array}{cccc}
 \eta' & -(2 \mu+1) u' & \frac{3 \eta'}{2} & -3 \mu u' \\
 -(2 \mu-1) u' &-\eta' & 3 \mu u' & \frac{3 \eta'}{2} \\
 \frac{\eta'}{2} & \mu u' & 0 & -3 u' \\
 -\mu u' & \frac{\eta'}{2} & 3 u' & 0 \\
\end{array}
\right)
\end{equation}
and $B^0_3$ the matrix
\begin{equation}
    B^0_{3}=\frac{2 \pi B_0^{-3/2}}{\sqrt{1-\mu}^2}
    \left(
\begin{array}{cccc}
    -(2 \mu +3) \phi_{31}^c+3 \mu \phi_{33}^c\\
    (2 \mu-3) \phi_{31}^s-3 \mu \phi_{33}^s\\
    -\mu \phi_{31}^c+3 \phi_{33}^c\\
    - \mu \phi_{31}^s+3 \phi_{33}^s
\end{array}
\right).
\end{equation}
In order to diagonalize the system of equations in \cref{eq:systempsi03}, we introduce the transformation $\Psi^0_{3}=T_{3}\sigma^0_{3}$, with $\sigma^0_{3}=[\sigma^0_{31} {,} \sigma^0_{32}{,} \sigma^0_{33} {,}\sigma^0_{34}]^T$ and $T_{3}$ the matrix
\begin{equation}
    T_{3}= e^{-3\eta/2}\left(
\begin{array}{cccc}
 e^{2 \eta}-1 & 2 i e^{2 \eta} \sinh (\eta) & -3 e^{2 \eta}-1 & -i e^{\eta} \left(e^{2 \eta}+3\right) \\
 e^{2 \eta}-1 & -2 i e^{2 \eta} \sinh (\eta) & -3 e^{2 \eta}-1 & i e^{\eta} \left(e^{2 \eta}+3\right) \\
 e^{2 \eta}+3 & -i e^{\eta} \left(3 e^{2 \eta}+1\right) & 3-3 e^{2 \eta} & 6 i e^{2 \eta} \sinh (\eta) \\
 e^{2 \eta}+3 & i e^{\eta} \left(3 e^{2 \eta}+1\right) & 3-3 e^{2 \eta} & -6 i e^{2 \eta} \sinh (\eta) \\
\end{array}
\right)
\end{equation}
The quantities $\sigma^0_{3}$ then satisfy
\begin{equation}
    \sigma^{0'}_{3}-\left(
\begin{array}{cccc}
\frac{i u'}{\cosh \eta} & 0 & 0 & 0 \\
0 & 3\frac{i u'}{\cosh \eta} & 0 & 0 \\
0 & 0 & -\frac{i u'}{\cosh \eta} & 0 \\
0 & 0 & 0 &-3 \frac{i u'}{\cosh \eta}
\end{array}
\right)\sigma^0_{3}=F^0_{3},
\label{eq:sigma03eqs}
\end{equation}
with $F^0_{3}=T_{3}^{-1} B^0_{3}$.

The decoupling of the $n=4$ case in \cref{eq:psi0n} can be performed in an analogous manner, where the equation for $\psi^0_{4}=\psi^0_{40}+ \psi_{42}^{0c} \cos 2 u + \psi_{42}^{0s} \sin 2u+\psi_{44}^{0c} \cos 4u+\psi_{44}^{0s} \sin 4 u$ can be written as
\begin{equation}
    \Psi^{0'}_{4}=A^0_{4} \Psi^0_{4}+B^0_{4},
\label{eq:systempsi04}
\end{equation}
with $\Psi^0_{4}=B_0^{-4/2}[\psi_{040}{,} \psi_{042}^c  {,}\psi_{042}^s{,} \psi_{044}^c {,}\psi_{044}^s]^T$ and $A^0_{4}$ the matrix
\begin{equation}
    A^0_{4}= \left(
\begin{array}{ccccc}
 0 & -3 \mu u' & \frac{3 \eta'}{2} & 0 & 0 \\
 -4 \mu u' & 0 & 2 u' & 2 \eta' & 4 \mu u' \\
 2 \eta' & -2 u' & 0 & -4 \mu u' & 2 \eta' \\
 0 & \frac{\eta'}{2} & -\mu u' & 0 & 4 u' \\
 0 & \mu u' & \frac{\eta'}{2} & -4 u' & 0 \\
\end{array}
\right)
\end{equation}
The diagonalizing matrix $T_4$ is given by
\begin{equation}
    T_{4}= \frac{1}{64}\left(
\begin{array}{ccccc}
 -6 \sinh ^2\eta & -6 \sinh ^2\eta & 12 \sinh 2\eta & 12 \sinh 2\eta & 6 \cosh 2\eta+2 \\
 8 i \sinh \eta & -8 i \sinh \eta & -16 i \cosh \eta & 16 i \cosh \eta & 0 \\
 -4 \sinh 2\eta & -4 \sinh 2\eta & 16 \cosh 2\eta & 16 \cosh 2\eta & 8 \sinh 2\eta \\
 4 i \cosh \eta & -4 i \cosh \eta & -8 i \sinh \eta & 8 i \sinh \eta & 0 \\
 -\cosh 2\eta-3 & -\cosh 2\eta-3 & 4 \sinh 2\eta & 4 \sinh 2\eta & 4 \sinh ^2\eta \\
\end{array}
\right),
\end{equation}
and the functions $\sigma^0_{4}$ satisfy the following set of equations
\begin{equation}
    \sigma^{0'}_{4}-\left(
\begin{array}{ccccc}
4\frac{i u'}{\cosh \eta} & 0 & 0 & 0 & 0 \\
0 & -4\frac{i u'}{\cosh \eta} & 0 & 0 & 0 \\
0 & 0 & 2\frac{i u'}{\cosh \eta} & 0 & 0 \\
0 & 0 & 0 &2 \frac{i u'}{\cosh \eta} & 0 \\
0 & 0 & 0 & 0 & 0
\end{array}
\right)\sigma^0_{4}=F^0_{4}.
\label{eq:sigma04eqs}
\end{equation}

\section{System of Equations for the Vacuum Flux Surface Function at Arbitrary Order}
\label{app:generalpsi}

We now solve for $\psi^0_{n}$ to obtain a set of differential equations of the form
\begin{equation}
    \Psi^{0'}_{n}=A^0_n \Psi^0_n + B^0_n,
\end{equation}
where the components of $\psi^0_{n}$ are written as $\Psi^0_{n}=B_0^{-n/2}[\psi_{0n0} {,} \Psi_{n2}^{0c}  {,} \Psi_{n2}^{0s} ...]^T$ for $n$ even and $\Psi^0_{n}=B_0^{-n/2}[\Psi_{n1}^{0c}  {,} \Psi_{n1}^{0s}  {,} \Psi_{n3}^{0c} {,}  \Psi_{n3}^{0s} ...]^T$ for $n$ odd
Plugging the expansion of \cref{eq:analyticity} in \cref{eq:psi0n}, we find for $\Psi^0_{nm}=B_0^{-n/2}\psi^0_{nm}$ and $0\le m \le 2$
\begin{align}
    \Psi_{n0}^{0c'}&=(n+2)\left[\frac{\eta'}{4}\Psi_{n2}^{0c}-\frac{\mu u'}{2}\Psi_{n2}^{0s}\right]+B_0^{-n/2-1}F_{n0}^{0c}\\
    \Psi_{n1}^{0c'}&=-u'\left[\Psi_{n1}^{0s}+\frac{\mu}{2}(n+1)\Psi_{n1}^{0s}+\frac{\mu}{2}(n+3)\Psi_{n3}^{0s}\right]\nonumber\\
    &+\frac{\eta'}{4}\left[(n+1)\Psi_{n1}^{0c}+(n+3)\Psi_{n3}^{0c}\right]+B_0^{-n/2-1}F_{n1}^{0c},\\
    \Psi_{n1}^{0s'}&=-u'\left[\Psi_{n1}^{0s}+\frac{\mu}{2}(n+1)\Psi_{n1}^{0s}+\frac{\mu}{2}(n+3)\Psi_{n3}^{0s}\right]\nonumber\\
    &+\frac{\eta'}{4}\left[(n+1)\Psi_{n1}^{0c}+(n+3)\Psi_{n3}^{0c}\right]+B_0^{-n/2-1}F_{n1}^{0c},\\
    \Psi_{n2}^{0c'}&=n \frac{\eta'}{2}\Psi_{n0}^{0c}-2u'\Psi_{n2}^{0s}+\frac{n+4}{2}\left(\mu u' \Psi_{n4}^{0s}-\frac{\eta'}{2}\Psi_{n4}^{0c}\right)+B_0^{-n/2-1}F_{n2}^{0c}\\
    \Psi_{n2}^{0s'}&=2 u'\Psi_{0n2}^{c}-n \mu u' \Psi_{n0}^{0c}+\frac{n+4}{2}\left(\mu u' \Psi_{n4}^{0c}+\frac{\eta'}{2}\Psi_{n4}^{0s}\right)+B_0^{-n/2-1}F_{n2}^{0s},
\end{align}
and for $m\ge 3$
\begin{align}
    \Psi_{nm}^{0c'}&=\frac{\eta'}{4}\left[\Psi_{nm-2}^{0c}(n+2-m)+\Psi_{nm+2}^{0c}(n+2+m)\right]\nonumber\\
    &-u'\left[m \Psi_{nm}^{0s}-\frac{\mu}{2}\Psi_{nm-2}^{0s}(n+2-m)+\frac{\mu}{2}\Psi_{nm+2}^{0s}(n+2+m)\right]+B_0^{-n/2-1}F^{0c}_{nm}\\
    \Psi_{nm}^{0s'}&=\frac{\eta'}{4}\left[\Psi_{nm-2}^{0s}(n-2+m)+\Psi_{nm+2}^{0s}(n+2+m)\right]\nonumber\\
    &-u'\left[-m \Psi_{nm}^{0c}+\frac{\mu}{2}\Psi_{nm-2}^{0c}(n+2-m)-\frac{\mu}{2}\Psi_{nm+2}^{0c}(n+2+m)\right]+B_0^{-n/2-1}F^{0s}_{nm},
\end{align}
with $F^{0s}_{nm}$ and $F^{0c}_{nm}$ the $\sin m u$ and $\cos m u$ Fourier coefficients of $F^0_{n}$, respectively.

\bibliographystyle{jpp}
\bibliography{library}

\end{document}